\documentclass[article,aps,preprint,floatfix,nofootinbib,showpacs]{revtex4-1}
\pdfoutput=1
\usepackage{textpos} 
\usepackage{dcolumn}
\usepackage{bm}
\usepackage{graphicx}
\usepackage{amssymb,amsmath}
\usepackage{multirow}
\usepackage{units,changes}
\usepackage{color,url}
\usepackage{tabu}
\usepackage{array}
\usepackage{booktabs}
\usepackage[colorlinks=true,urlcolor=blue,anchorcolor=blue
,citecolor=blue,filecolor=blue,linkcolor=blue,menucolor=blue
,linktocpage=true,pdfproducer=medialab,pdfa=true]{hyperref}
\usepackage{colordvi}
\usepackage{subcaption}
\def\beqa{\begin{eqnarray}}
\def\eeqa{\end{eqnarray}}
\newcommand{\ov}{\overline}
\usepackage{soul}
\usepackage{bigstrut}

\begin{document} 
\preprint{$\begin{gathered}\includegraphics[width=0.1\textwidth]{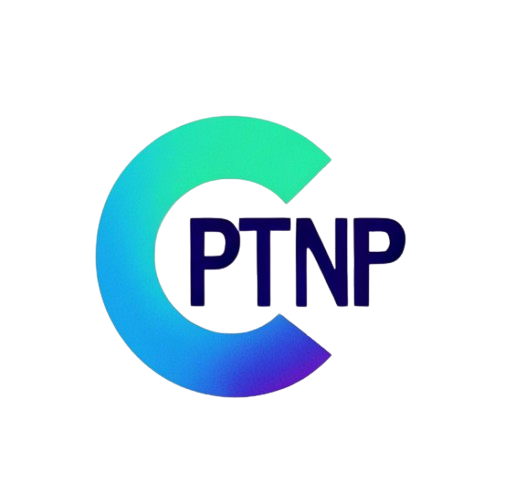}\end{gathered}$\,CPTNP-2025-006}
\begin{textblock}{12.5}(0.01,0.01)
  \raggedleft 
  {\text{}}  
\end{textblock}

\title{ Searching for elusive dark Higgs boson in spin-1/2 \\
inelastic dark matter models at Belle II } 
\def\slash#1{#1\!\!\!/}

\author{P. Ko$^{1,2}$, Youngjoon Kwon$^3$, Chih-Ting Lu$^{4,5}$, Xinqi Wei$^4$}
\affiliation{$^1$School of Physics, KIAS, Seoul 02455, Korea}
\affiliation{$^2$ Quantum Universe Center (QUC), KIAS, Seoul 02455, Korea}
\affiliation{$^3$ Department of Physics, Yonsei University, Seoul 03722, Republic of Korea}
\affiliation{$^4$ Department of Physics and Institute of Theoretical Physics, Nanjing Normal University, Nanjing, 210023, China}
\affiliation{$^5$ Nanjing Key Laboratory of Particle Physics and Astrophysics, Nanjing, 210023, China}


\begin{abstract} 
Spin-1/2 inelastic dark matter (DM) models are popular among sub-GeV to GeV 
thermal DM scenarios due to the dominant role of co-annihilation in determining the DM relic abundance. In these models, the dark Higgs boson plays a crucial role in generating the mass of the new gauge boson, the dark photon ($A'$), and in establishing the mass splitting between the excited ($\chi_2$) and ground ($\chi_1$) states of DM. 
In particular, the Compton scattering $\chi_1 A' \rightarrow \chi_2^* \rightarrow \chi_1 A'$ 
and its $t$-channel crossed process, $\chi_1 \chi_1 \rightarrow A' A'$, remain unitary 
for high energy longitudunal dark photon, only if the contribution of the dark Higgs boson 
is included. 
However, experimental searches for the dark Higgs boson have received relatively little attention. In particular, when the dark Higgs boson mass exceeds twice that of the DM excited state, its decay signatures become semi-visible or invisible, making detection challenging with current light scalar search strategies. 
In this work, we explore the prospects for detecting the elusive dark Higgs boson in spin-1/2 inelastic DM models at Belle II via dark Higgs-strahlung and rare $B$ meson decay processes. Our analysis indicates that both the inclusive signature of two displaced dilepton vertices and the additional missing energy from dark Higgs boson decays serve as robust indicators of its presence. Furthermore, we assess the future potential for detecting the dark Higgs boson with the proposed far detector related to Belle II, GAZELLE.
\end{abstract}

\maketitle

\section{Introduction}

The particle nature of dark matter (DM) remains one of the most significant unknowns in particle physics, astronomy, and cosmology. Most evidence for the existence of DM comes from its gravitational interactions~\cite{Trimble:1987ee,Barack:2018yly}. The mass range of DM spans from nearly massless particles to those on the order of a solar mass~\cite{Battaglieri:2017aum,Lin:2019uvt,Ferreira:2020fam,Belenchia:2021rfb}. Similarly, the coupling strength between DM and Standard Model (SM) particles varies widely across different DM models~\cite{PerezdelosHeros:2020qyt}. Searching for DM is very challenging without assumptions to narrow down both the DM mass and coupling strength ranges. Thermally produced DM through the freeze-out mechanism is an attractive approach, as it implies a significant non-gravitational interaction between DM and SM particles and suggests a DM mass range from keV to PeV~\cite{Kolb:1990vq,Berlin:2017ftj,Frumkin:2022ror}.

Weakly Interacting Massive Particles (WIMPs) are well-motivated thermal DM candidates, with masses ranging from a few GeV to tens of TeV and interaction strengths with SM particles similar to weak interactions~\cite{Arcadi:2017kky,Arcadi:2024ukq}. There are three typical ways to search for WIMPs: direct detection~\cite{Schumann:2019eaa}, indirect detection~\cite{Gaskins:2016cha}, and collider~\cite{Kahlhoefer:2017dnp,Boveia:2018yeb} experiments. However, no concrete evidence has been found from any of these searches thus far. Consequently, researchers have extended their focus from WIMPs to more general hidden sector DM models, which allow for a wider range of DM masses and interaction strengths~\cite{Battaglieri:2017aum}. Among various hidden sector DM models, sub-GeV DM models have gained popularity as new thermal DM freeze-out processes~\cite{Hochberg:2014dra,DAgnolo:2015ujb,Kuflik:2015isi,Cline:2017tka,DAgnolo:2017dbv,Frumkin:2021zng} and search strategies~\cite{Knapen:2017xzo,Lin:2019uvt} continue to emerge.

Inelastic DM models are a type of thermal DM model with masses ranging from sub-GeV to the electroweak scale, where the DM relic abundance is primarily contributed by co-annihilation~\cite{TuckerSmith:2001hy,Baek:2014kna,Izaguirre:2015zva,DEramo:2016gqz,Izaguirre:2017bqb,Berlin:2018jbm,Mohlabeng:2019vrz,Tsai:2019buq,Okada:2019sbb,Ko:2019wxq,Duerr:2019dmv,Duerr:2020muu,Ema:2020fit,Kang:2021oes,Batell:2021ooj,Guo:2021vpb,Li:2021rzt,Filimonova:2022pkj,Bertuzzo:2022ozu,Mongillo:2023hbs,CarrilloGonzalez:2021lxm,Lu:2023cet,Berlin:2023qco,Heeba:2023bik,Garcia:2024uwf,Liu:2025abt}. 
Due to this co-annihilation process, cosmological constraints from the Cosmic Microwave Background (CMB) as well as direct and indirect detection constraints are relaxed. However, the cosmological constraint from Big Bang Nucleosynthesis (BBN) still requires the DM mass to be larger than about $10$ MeV~\cite{Depta:2019lbe,Sabti:2019mhn}. Consequently, accelerator experiments become a powerful way to explore inelastic DM models~\cite{Izaguirre:2015zva,Izaguirre:2017bqb,Berlin:2018jbm}. In this work, we focus on fermionic inelastic DM models with $U(1)_D$ dark gauge symmetry~\cite{Ko:2019wxq,Kang:2021oes,Lu:2023cet}. The DM candidate is a Dirac fermion field that splits into two Majorana fermion fields after the spontaneous symmetry breaking (SSB) of $U(1)_D$, hence it often called pseudo-Dirac DM\footnote{In inelastic DM models, the DM candidate can also be the complex scalar field which will split into two real scalar fields after the SSB of $U(1)_D$~\cite{Baek:2014kna,Baek:2020owl,Kang:2021oes}.}.

Regarding the mass splitting between the excited and the ground states in the 
spin-1/2 inelastic DM models, this mass splitting ($\Delta_\chi \equiv M_{\chi_2} - M_{\chi_1}$) is given by hand, by introducing a dim-3 operator that breaks underlying gauge symmetry explicitly and softly for spin-1/2 inelastic DM case (for example, see Ref.~\cite{TuckerSmith:2001hy}):
\[
\Delta_\chi \overline{\chi^c} \chi + H.c..
\]
Likewise, the mass splitting in the scalar inelastic DM models is generated by assuming 
$U(1)_D$ symmetry is explicitly and softly broken by a dim-2 operator~\cite{TuckerSmith:2001hy}. 
In either case, there is no dark Higgs involved in generating the mass splitting of 
inelastic DM models, and it is generated by hand from soft gauge symmetry breaking dim-2 
or dim-3 operators for scalar and spin-1/2 inelastic DM cases.

However, in such models, processes involving inelastic DM and a massive dark photon, such as Compton-like scattering $\chi_1 A'_L \rightarrow \chi_2^* \rightarrow \chi_1 A'_L$, and the $t$-channel crossed process $\chi_1 \chi_1 \rightarrow A'_L A'_L$, where $A'_L$ denotes the longitudinal component of the dark photon, can lead to unitarity violation. This issue arises because the scattering amplitudes contain terms that grow with energy, resulting in a breakdown of unitarity at high energies~\cite{Ko:2019wxq}. Introducing a dark Higgs boson resolves this problem by canceling the problematic energy-growing terms, thereby restoring unitarity and ensuring the theoretical consistency of spin-1/2 inelastic DM models~\cite{Ko:2019wxq}\footnote{See Appendix A in Ref.~\cite{Ko:2019wxq} for more details on the unitarity issue in $\chi_1 \chi_1 \rightarrow A'_L A'_L$.}.

In summary, in the inelastic DM models with dark gauge symmetry $U(1)_D$ spontaneously broken into its subgroup $Z_2$, the dark Higgs mechanism not only generates the mass of dark photon but also induces the mass splitting between the two Majorana fermions. The lighter one (ground state) is the DM candidate while the heavier one (excited state) can decay into the ground state plus SM particles if the mass splitting is large enough. Because $U(1)_D$ is broken into its $Z_2$ subgroup, the DM ground state is stable. 
Additionally, due to the kinematic mixing between $U(1)_D$ and the SM $U(1)_Y$ gauge bosons, the dark photon can further interact with SM fermion fields. Consequently, the DM ground state and excited state can co-annihilate into SM particles via the $s$-channel dark photon exchange. Although the dark Higgs boson is not directly involved in this co-annihilation process, it plays a crucial role in generating both the mass splitting between the excited and ground states of DM and the mass of dark photon.

The decay of light scalars into visible~\cite{Aubert:2009cp,Aubert:2009cka,Lees:2011wb,Lees:2012iw,LHCb:2020ysn,Belle:2021rcl,Cheung:2024oxh} and/or invisible~\cite{delAmoSanchez:2010ac,Seong:2018gut,Ablikim:2020qtn} final states has been widely studied. However, in inelastic DM models, the signatures of dark Higgs boson decays become semi-visible or invisible when the mass of the dark Higgs boson is larger than twice the mass of the DM excited state. Semi-visible decays refer to final states that include both visible SM particles and the invisible component from DM ground state. This type of semi-visible dark Higgs boson decay has received little attention~\cite{Li:2021rzt,Ferber:2023iso}. If the DM excited state is a long-lived particle (LLP), the decay of the dark Higgs boson can produce a novel signature with two dilepton displaced vertices. We focus on two specific dark Higgs boson production processes at Belle II: dark Higgs-strahlung and 
$B$ meson rare decay processes. Identifying the inclusive two dilepton displaced vertices from these processes would provide important evidence for the elusive dark Higgs boson. Additionally, we study how the invisible dark Higgs boson decay modifies previous searches for inelastic DM that did not involve a dark Higgs boson and the $B\to K\nu\bar\nu$ measurement~\cite{Belle-II:2023esi}. The signatures of extra missing energy from these processes can also help indicate the existence of the dark Higgs boson in inelastic DM models. Finally, in certain regions of the parameter space, the DM excited state can become very long-lived. To detect this DM excited state from dark Higgs boson decays, we further examine the potential for searching for such signatures using the proposed new far detector related to Belle II, GAZELLE~\cite{Dreyer:2021aqd}, in this study.

The remainder of this paper is organized as follows. In Sec.~\ref{sec2}, we review fermionic inelastic DM models. Sec.~\ref{sec3} examines relevant constraints on the model parameters. Our main focus, searching for the semivisible dark Higgs boson at Belle II, is presented in Sec.~\ref{sec4}. In Sec.~\ref{sec5}, we attempt to explain the excess observed in $B\to K\nu\overline{\nu}$ measurements at Belle II through dark Higgs boson decays and explore the future potential for detecting the dark Higgs boson with the proposed GAZELLE detector. Finally, we conclude our findings in Sec.~\ref{sec6}.

\section{Spin-1/2 inelastic dark matter models} 
\label{sec2}

\subsection{The model} 

This section reviews spin-1/2 inelastic DM models with $U(1)_D$ dark gauge symmetry that is spontaneously broken into its $Z_2$ subgroup. 
We consider a dark sector with a singlet complex scalar $\Phi$ and a Dirac fermion $\chi$ as the dark Higgs and DM fields, respectively. 
Both $\Phi$ and $\chi$ are charged under $U(1)_D$ but neutral with respect to the SM gauge symmetry. 
The $U(1)_D$ charges for $\Phi$ and $\chi$ fields are assigned as $ Q_D(\Phi)= +2 $ and $ Q_D(\chi)= +1 $, for which there appears new gauge invariant operator, $\overline{\chi^c} \chi \Phi^{\ast}$ 
\footnote{See the last term in Eq.~(\ref{eq:LDMf}), which breaks the $U(1)_D$ symmetry down to its $Z_2$ subgroup.
A similar setup can be considered for complex scalar DM $X$, which gives rise to inelastic real scalar DM after the $U(1)_D$ symmetry is broken~\cite{Baek:2014kna,Baek:2020owl,Kang:2021oes}.
}. 
The SM-like Higgs doublet and other SM particles do not carry $U(1)_D$ charges. After the SSB of the $U(1)_D$, an accidental residual $Z_2$ symmetry, $\chi_1\rightarrow -\chi_1$, is expected to remain, making $\chi_1$ stable and a DM candidate in our universe.

The scalar part of renormalizable and gauge invariant Lagrangian density is
\begin{equation}
{\cal L}_{\text{scalar}} = 
| D_{\mu}H |^2 + | D_{\mu}\Phi |^2 - V(H,\Phi), 
\label{eq:Lf}
\end{equation}
with
\begin{equation}
D_{\mu}H = (\partial_\mu +i\frac{g}{2}\sigma_a W^a_{\mu}+i\frac{g'}{2}B_{\mu})H,
\label{eq:kin1}
\end{equation}
\begin{equation}
D_{\mu}\Phi = (\partial_\mu +ig_D Q_D(\Phi) X_{\mu})\Phi,
\label{eq:kin2}
\end{equation}
where $ W^a_{\mu} $, $ B_{\mu} $, and $ X_{\mu} $ are gauge potentials of the $SU(2)_L$, $U(1)_Y$ and $U(1)_D$ with gauge couplings $g$, $g'$ and $g_D$, respectively. 
The $ \sigma_a $ is the Pauli matrix and $ a $ runs from 1 to 3. The scalar potential in Eq.~(\ref{eq:Lf}) is given by
\begin{align}
V(H,\Phi) = & 
-\mu^2_H H^{\dagger}H +\lambda_H (H^{\dagger}H)^2 -\mu^2_\Phi \Phi^{\ast}\Phi +\lambda_\Phi (\Phi^{\ast}\Phi)^2 
\nonumber  \\ &
+\lambda_{H\Phi}(H^{\dagger}H)(\Phi^{\ast}\Phi),
\label{eq:Vs}
\end{align} 
where all coefficients are assumed to be real. We then expand $H,\Phi$ fields around the vacuum expectation value (VEV) with the unitary gauge,
\begin{equation}
H(x) = \frac{1}{\sqrt 2} 
\left(
\begin{tabular}{c}
0
\\
$v + h(x)$
\end{tabular}
\right)
\;\;\; , \;\;\; 
\Phi (x) = \frac{1}{\sqrt 2} \left( v_D + h_D (x) \right). 
\label{eq:expand}
\end{equation} 
The interaction states $ (h, h_D) $ would be rotated to the mass states $ (h_1, h_2) $ 
via the mixing angle $ \theta $ as 
\begin{equation}
\begin{pmatrix} h \\\ h_D \end{pmatrix} =
\begin{pmatrix}
\cos\theta && \sin\theta \\
-\sin\theta && \cos\theta
\end{pmatrix}
\begin{pmatrix}
h_1 \\ h_2
\end{pmatrix}\,,
\label{eq:mix}
\end{equation} 
where we assume $ m_{h_1} > m_{h_2} $, such that $ h_1 $ is assigned as SM-like Higgs boson with $ m_{h_1} = 125 $ GeV and $h_2$ is the light dark Higgs boson.

The Lagrangian density of DM sector is 
\begin{equation}
{\cal L}_{\chi} = 
\ov{\chi}(i{\rlap{\,/}\partial}+g_D{\rlap{\,/}X}-M_{\chi})\chi -(\frac{f}{2}\ov{\chi^c}\chi\Phi^{\ast}+H.c.),
\label{eq:LDMf}
\end{equation}
where $f$ is assumed to be a real parameter. Here, parity conservation is assumed for simplicity, so that the $\chi_L$ and $\chi_R$ fields couple identically to the $\Phi$ field. A more general setup that relaxes this assumption can be found in Refs.~\cite{Izaguirre:2015zva,Duerr:2019dmv,Garcia:2024uwf}. In order to decompose the Dirac fermion $\chi$ into a pair of two independent Majorana fermions, 
$\chi_1$ and $\chi_2$, we set  
\begin{equation}
\chi_{1,2}=\frac{1}{\sqrt{2}}(\chi(x)\mp\chi^{c}(x)).
\end{equation}
After expanding $H$, $\Phi$ fields around the VEV with the unitary gauge as shown in Eq.~(\ref{eq:expand}),
the Eq.~(\ref{eq:LDMf}) can be written as 
\begin{align}
{\cal L}_{\chi} = & 
\frac{1}{2}\ov{\chi_2}(i{\rlap{\,/}\partial}-M_{\chi_2})\chi_2 +\frac{1}{2}\ov{\chi_1}(i{\rlap{\,/}\partial}-M_{\chi_1})\chi_1  
\nonumber  \\ &
-i\frac{g_D}{2}(\ov{\chi_2}{\rlap{\,/}X}\chi_1 -\ov{\chi_1}{\rlap{\,/}X}\chi_2) -\frac{f}{2}h_D (\ov{\chi_2}\chi_2 -\ov{\chi_1}\chi_1)
\label{eq:LDMf2}
\end{align}
where $\chi_1$, $\chi_2$ masses and their mass splitting can be represented as
\begin{equation}
M_{\chi_{1,2}} = M_{\chi}\mp fv_D,
\end{equation}
and
\begin{equation}
\Delta_{\chi}\equiv (M_{\chi_2} - M_{\chi_1}) = 2fv_D.
\end{equation} 
Furthermore, according to Eq.~(\ref{eq:mix}) and Eq.~(\ref{eq:LDMf2}), the Lagrangian for the interactions of $ h_{1,2} $ with SM fermion pairs and DM excited/ground state pairs is given by 
\begin{align}
{\cal L}_{\text{int}} \supset & 
-(\cos\theta h_1 + \sin\theta h_2)\sum_{f}\frac{m_f}{v}\overline{f}f
\nonumber  \\ &
+\dfrac{f}{2}(\sin\theta h_1 - \cos\theta h_2)(\overline{\chi_2}\chi_2 -\overline{\chi_1}\chi_1).
\end{align}

Finally, since all SM fermions don't carry $U(1)_D$ charges, the only way for the new $ X_\mu $ 
boson and SM fermions to interact is via the kinetic mixing between $ B_{\mu\nu} $ and 
$ X_{\mu\nu} $. The Lagrangian density of this part can be represented as
\begin{equation}
{\cal L}_{X,\text{gauge}} = -\frac{1}{4}  X_{\mu\nu}X^{\mu\nu} 
-\frac{\sin\epsilon}{2} B_{\mu\nu}X^{\mu\nu}, 
\label{eq:Zps}
\end{equation}
where $\epsilon$ is the kinetic mixing parameter between these two $U(1)$s. 
If we apply the linear order approximation in $\epsilon$, the extra interaction terms for SM fermions and $A^{\prime}$ can be written as
\begin{equation}
{\cal L}_{A^{\prime}f\overline{f}} = -\epsilon e c_W \sum_{f} x_f \ov{f} {\rlap{\,/}A^{\prime}} f
\label{eq:Zpffbar}
\end{equation}
where $ c_W $ is the weak mixing angle and $ x_l = -1 $, $ x_{\nu} = 0 $, $ x_q = \frac{2}{3} $ or $ \frac{-1}{3} $ depending on the electrical charge of quark.
The dark photon mass can be approximated as
\begin{equation}
m_{A^{\prime}} \simeq g_D Q_D(\Phi) v_D.
\label{eq:Zpmass}
\end{equation}
Notice the correction from the kinetic mixing term is second order in $ \epsilon $ which can be safely neglected here.

\subsection{The relevant partial decay widths of particles in inelastic DM models}

The partial decay widths of SM-like Higgs boson $h_1$ in this model are written as  
\begin{equation}
\Gamma_{h_1} = \cos^2\theta\Gamma_{h_{\text{SM}}} + \sin^2\theta \left[\widehat{\Gamma}(h_1\rightarrow\chi_{1,2}\chi_{1,2})+\widehat{\Gamma}(h_1\rightarrow A^{\prime}A^{\prime})\right]+\Gamma (h_1\rightarrow h_2 h_2)
\end{equation} 
and 
\begin{align} 
& \widehat{\Gamma}(h_1\rightarrow\chi_{1,2}\chi_{1,2}) = \frac{f^2}{16\pi m^2_{h_1}}(m^2_{h_1}-4M^2_{\chi_{1,2}})^{3/2},
\nonumber  \\ &
\widehat{\Gamma}(h_1\rightarrow A^{\prime}A^{\prime}) = \frac{g^2_D m^2_{A^{\prime}}}{2\pi m_{h_1}}\sqrt{1-\frac{4m^2_{A^{\prime}}}{m^2_{h_1}}}(3-\frac{m^2_{h_1}}{m^2_{A^{\prime}}}+\frac{m^4_{h_1}}{4m^4_{A^{\prime}}}), 
\nonumber  \\ & 
\Gamma (h_1\rightarrow h_2 h_2) = \frac{\lambda^2_{h_1 h_2 h_2}}{32\pi m_{h_1}}\sqrt{1-\frac{4m^2_{h_2}}{m^2_{h_1}}},
\end{align}
where 
\begin{equation}
\lambda_{h_1 h_2 h_2} = \frac{(m^2_{h_1}+2m^2_{h_2})(v\cos\theta -v_D \sin\theta)\sin2\theta}{2vv_D}.
\end{equation} 
The $ \Gamma_{h_{\text{SM}}} $ is the total width of Higgs boson in the SM with a theoretical value of $4.03$ MeV~\cite{LHCHiggsCrossSectionWorkingGroup:2011wcg}. 

\begin{figure}[t!]\centering
\includegraphics[width=0.48\textwidth]{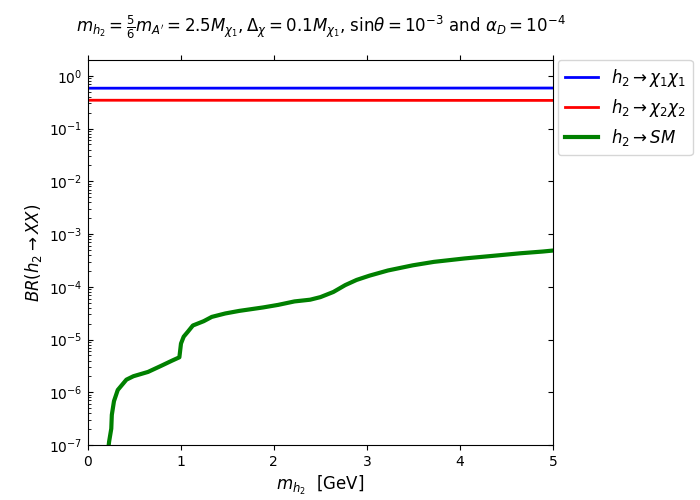}
\includegraphics[width=0.48\textwidth]{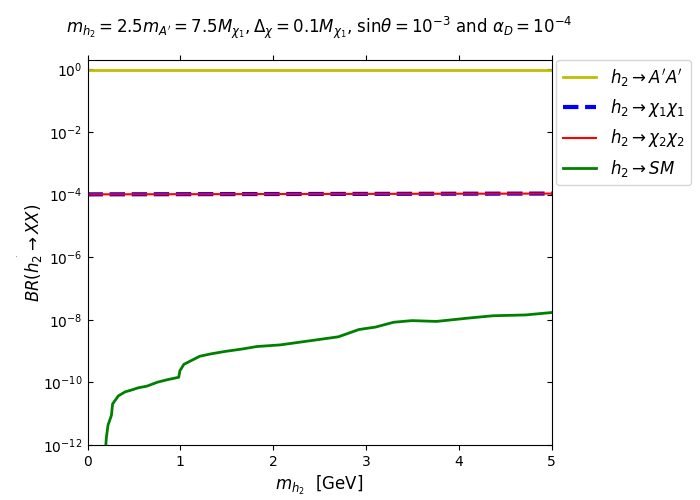}
\caption{ 
The decay branching ratio of $h_2$. Two benchmark points: $ m_{h_2} = \frac{5}{6}m_{A^{\prime}} = 2.5 M_{\chi_1} $ (left panel) and $ m_{h_2} = 2.5 m_{A^{\prime}} = 7.5 M_{\chi_1} $ (right panel). Here we fix $\Delta_{\chi} = 0.1 M_{\chi_1}$, $ \sin\theta = 10^{-3} $ and $ \alpha_D = 10^{-4} $.
}
\label{fig:h2 decay ratio}
\end{figure}

Similarly, the decay width of $ h_2 $ can be written as 
\begin{equation}
\Gamma_{h_2} = \sin^2\theta\widehat{\Gamma}(h_2\rightarrow\text{SM}) + \cos^2\theta\left[\widehat{\Gamma}(h_2\rightarrow\chi_{1,2}\chi_{1,2})+\widehat{\Gamma}(h_2\rightarrow A^{\prime}A^{\prime})\right],  
\label{eq:h2 decay eq.}
\end{equation} 
where $ \widehat{\Gamma}(h_2\rightarrow\text{SM}) $ includes the photon pair, charged lepton pairs and meson pairs (low $ m_{h_2} $) or gluon pair, quark pairs (high $ m_{h_2} $). We closely follow Appendix A in Ref.~\cite{Liu:2014cma} for these formulas and will not repeat them again here. The formulas for $\widehat{\Gamma}(h_2\rightarrow\chi_{1,2}\chi_{1,2})$ and $\widehat{\Gamma}(h_2\rightarrow A^{\prime}A^{\prime})$ are the same as those for $\widehat{\Gamma}(h_1\rightarrow\chi_{1,2}\chi_{1,2})$ and $\widehat{\Gamma}(h_1\rightarrow A^{\prime}A^{\prime})$, except with $m_{h_1}$ replaced by $m_{h_2}$.
In Fig.~\ref{fig:h2 decay ratio}, we consider the mass spectrum $ m_{h_2} = \frac{5}{6}m_{A^{\prime}} = 2.5 M_{\chi_1} $ on the left panel and $ m_{h_2} = 2.5 m_{A^{\prime}} = 7.5 M_{\chi_1} $ on the right panel for the decay branching ratio of $ h_2 $. Here we fix $\Delta_{\chi} = 0.1 M_{\chi_1}$, $ \sin\theta = 10^{-3} $ and $ \alpha_D = 10^{-4} $.
It is evident that when the decay $h_2\to A'A'$ is kinematically forbidden, the channel $h_2\to\chi_{1,2}\chi_{1,2}$ dominates, allowing us to safely neglect contributions from $h_2\to$SM. Conversely, once $h_2\to A'A'$ becomes kinematically accessible, it overwhelmingly dominates, and the contributions from both $h_2\to\chi_{1,2}\chi_{1,2}$ and $h_2\to$SM can be ignored. 

\begin{figure}[t!]\centering
\includegraphics[width=0.48\textwidth]{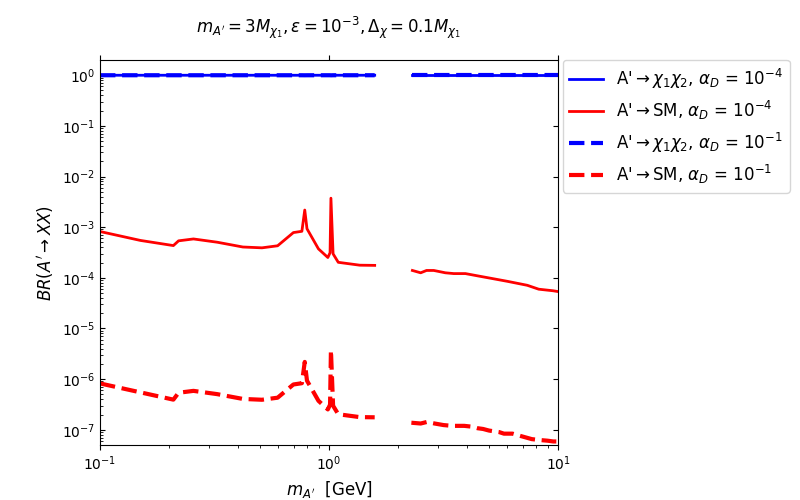}
\includegraphics[width=0.48\textwidth]{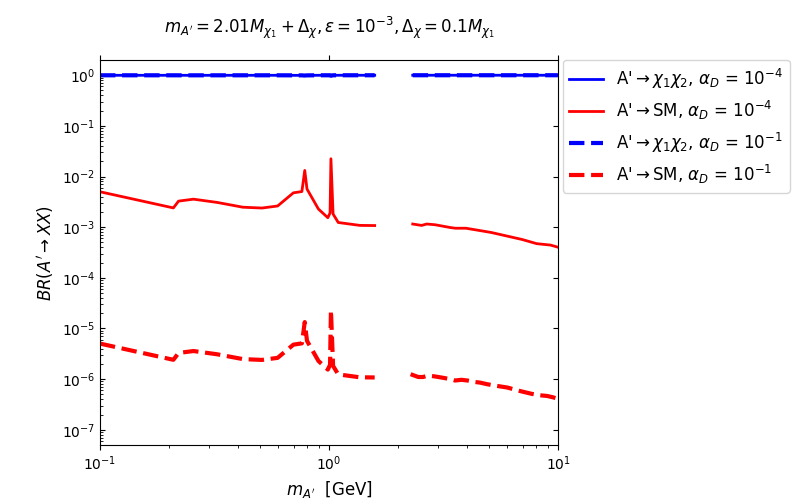}
\caption{ 
The decay branching ratio of $A^{\prime}$. Two benchmark points: $m_{A'}=3 M_{\chi_1}$ (left panel) and $m_{A'}=2.01 M_{\chi_1}+\Delta_{\chi}$ (right panel) are considered. Here we fix $\epsilon = 10^{-3}$, and $\Delta_{\chi} = 0.1 M_{\chi_1}$, but varying $\alpha_D = 10^{-4}$, and $10^{-1}$.
}
\label{fig:A' decay ratio}
\end{figure}

On the other hand, the decay width of $ A^{\prime} $ can be represented as 
\begin{equation}
\Gamma_{A^{\prime}} = \Gamma (A^{\prime}\rightarrow\chi_1\chi_2) +\Gamma (A^{\prime}\rightarrow f\overline{f})
\end{equation} 
and 
\begin{align} 
& \Gamma (A^{\prime}\rightarrow\chi_1\chi_2) = \frac{g^2_D}{24\pi m^3_{A^{\prime}}}\left[ 2m^2_{A^{\prime}}-M^2_{\chi_1}-M^2_{\chi_2}+6M_{\chi_1}M_{\chi_2}+
\frac{(M^2_{\chi_2}-M^2_{\chi_1})^2}{m^2_{A^{\prime}}} \right]
\nonumber  \\ &
\times\sqrt{\left[m^2_{A^{\prime}}-(M_{\chi_1}+M_{\chi_2})^2\right] \left[ m^2_{A^{\prime}}-(M_{\chi_1}-M_{\chi_2})^2\right]},
\nonumber  \\ &
\Gamma (A^{\prime}\rightarrow f\overline{f}) = \frac{\epsilon^2 e^2 C_f x^2_f}{12\pi}m_{A^{\prime}}(1+2\frac{m^2_f}{m^2_{A^{\prime}}})\sqrt{1-4\frac{m^2_f}{m^2_{A^{\prime}}}}, 
\label{eq:A' decay eq.}
\end{align} 
where $C_{\ell} = 1$ for charged leptons and $C_q = 3$ for quarks. We consider the mass spectrum $m_{A'}=3 M_{\chi_1}$ on the left panel and $m_{A'}=2.01 M_{\chi_1}+\Delta_{\chi}$ on the right panel for the decay branching ratios of $A'$ in Fig.~\ref{fig:A' decay ratio}. Here $\epsilon = 10^{-3}$, and $\Delta_{\chi} = 0.1 M_{\chi_1}$ are fixed, but varying $\alpha_D = 10^{-4}$, and $10^{-1}$. Note that for $m_{A^{\prime}} \lesssim 1.6$ GeV, $A^{\prime}$ predominantly decay into meson final states. Accordingly, we utilize the data from Ref.~\cite{Buschmann:2015awa} and apply the branching ratios for $A^{\prime}$ decaying into a muon pair to rescale its partial decay width into SM particles. Moreover, we observe that $A^{\prime}$ primarily decays into $\chi_1 \chi_2$.

\subsection{The target parameter space and decay width of $\chi_2$}

In this study, we focus on the scenario $ m_{A'} > M_{\chi_1} + M_{\chi_2} $ so that the decay $ A^{\prime}\rightarrow\chi_1\chi_2 $ is kinematically allowed and becomes the dominant mode, as shown in Fig.~\ref{fig:A' decay ratio}. 
Additionally, we consider the co-annihilation dominant channel for the DM relic density in the early universe, restricting ourselves to a compressed mass spectrum with $ \Delta_{\chi} < 0.5 M_{\chi_1} $ in inelastic DM models~\cite{Izaguirre:2015zva,Izaguirre:2017bqb,Duerr:2019dmv,Duerr:2020muu}. 
Finally, we conservatively assume $ M_{\chi_1} > 100 $ MeV to ensure that the parameter space is  not constrained by BBN and other low-energy experiments.

In our analysis, $ \chi_1 $ is the only DM candidate. If the mass splitting $ \Delta_{\chi} $ cannot be ignored compared with $ M_{\chi_1} $, $ \chi_2 $ is unstable and will decay to $ \chi_1 $ and SM fermion pairs via the off-shell $ A' $. The complete analytical formulas for the total width of the three-body decay of $ \chi_2 $ can be found in Appendix B of Ref.~\cite{Giudice:2017zke}. In the mass range of interest, $ \chi_2 $ has three types of decay modes\footnote{Since the light $ A' $ is an isosinglet, it can mix with the $ \omega $ meson and decay to three pions, $A'\rightarrow \pi^+ \pi^- \pi^0$, if kinematically allowed. However, as shown in Fig.~1 of Ref.~\cite{Ilten:2018crw}, the contributions from the $A'\rightarrow \pi^+ \pi^- \pi^0$ mode are significant only in the vicinity of $ m_{A'}\approx m_{\omega} = 0.782 $ GeV. Therefore, we do not specifically study this region in our analysis.}, namely $ \chi_2\rightarrow\chi_1 e^{+}e^{-}$, $\chi_1\mu^{+}\mu^{-}$ and $\chi_1\pi^{+}\pi^{-} $. In our numerical calculations, the total decay width of $\chi_2$ is automatically calculated using \textbf{Madgraph5 aMC@NLO}~\cite{Alwall:2014hca}.  
For the partial decay width of $\chi_2\rightarrow\chi_1\pi^{+}\pi^{-} $, we rescale the partial decay width of $\chi_2\rightarrow\chi_1\mu^{+}\mu^{-} $ using the measured $ R(s) $ values~\cite{Zyla:2020zbs}\footnote{Another approach to handling the hadronic decay $\chi_2 \to \chi_1 \pi^+ \pi^-$ is based on vector meson dominance (VMD) theory~\cite{Garcia:2024uwf}. However, the difference between these two methods has only a slight impact on our results.}.

Finally, we focus on the mass spectrum where $m_{h_2} > 2M_{\chi_2}$, as shown in Fig.~\ref{fig:h2 decay ratio}. This setting leads to several significant differences between our work and the previous dark Higgs boson searches at Belle II presented in Ref.~\cite{Duerr:2020muu}. First, in our mass spectrum, the DM relic density is determined solely by the co-annihilation process, whereas in Ref.~\cite{Duerr:2020muu}, the annihilation process $\chi_1\chi_1 \to h_2 h_2$ ($p$-wave) can also occur. Second, our scenario includes novel semi-visible and invisible decay channels of the dark Higgs boson, while in Ref.~\cite{Duerr:2020muu}, the long-lived dark Higgs boson can decay only visibly into a pair of SM particles. Third, due to the different signal signatures studied in our work and in Ref.~\cite{Duerr:2020muu}, we explore two distinct parameter regions: 
(1) large $\alpha_D$ with small $\sin\theta$, and
(2) small $\alpha_D$ with moderate $\sin\theta$. 
In contrast, Ref.~\cite{Duerr:2020muu} focuses on the case where both $\alpha_D$ and $\sin\theta$ are small.

\section{Constraints on model parameters}
\label{sec3}

\begin{figure}[t!]\centering
\includegraphics[width=0.48\textwidth]{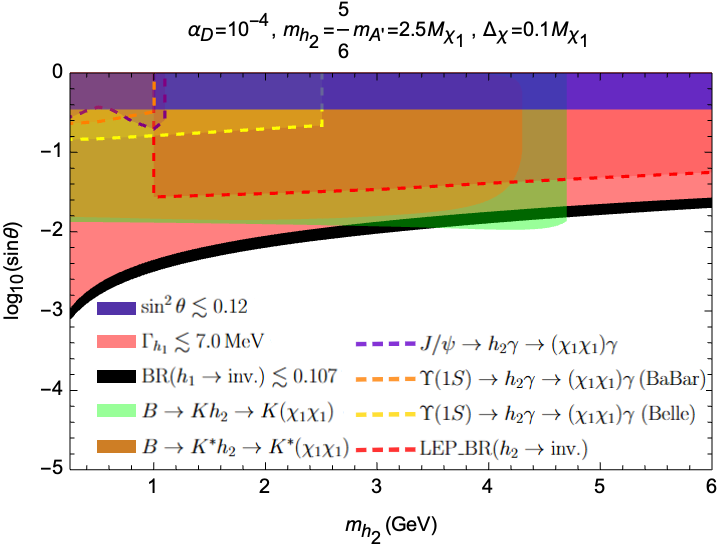}
\includegraphics[width=0.48\textwidth]{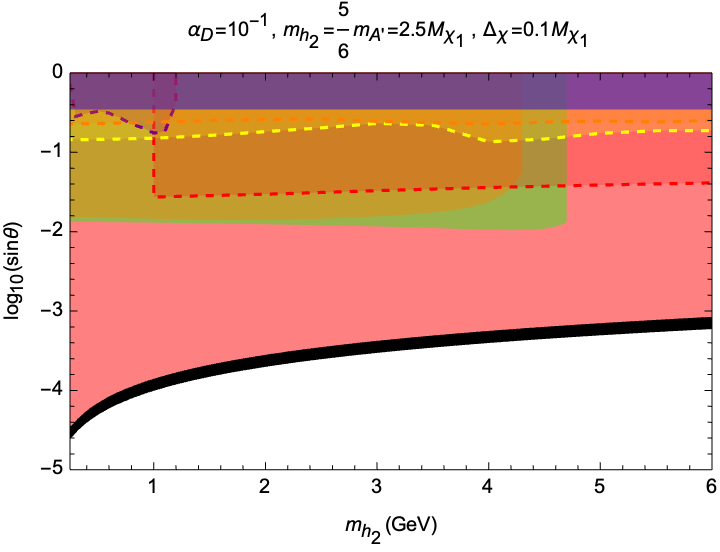}
\includegraphics[width=0.48\textwidth]{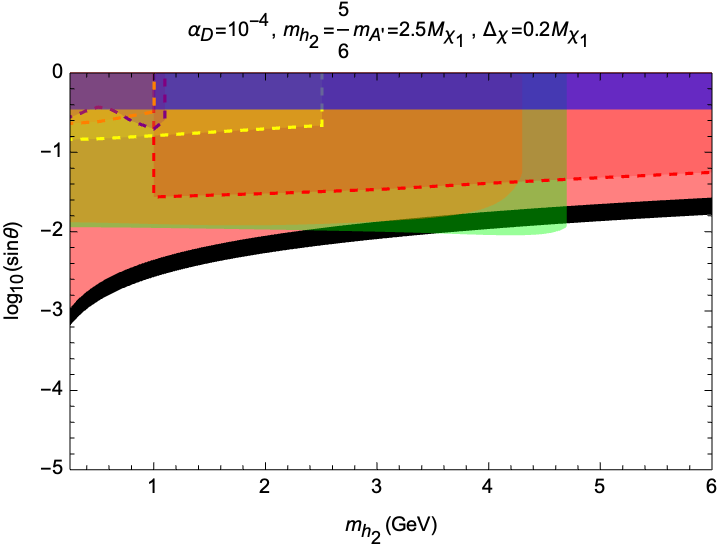}
\includegraphics[width=0.48\textwidth]{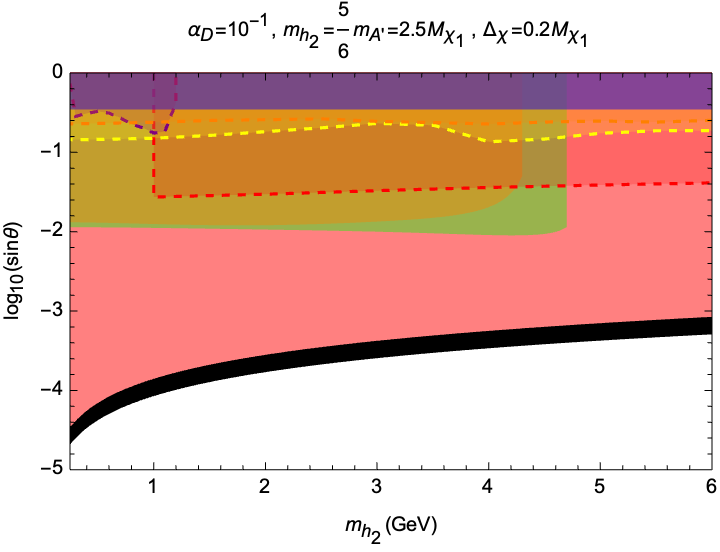}
\caption{ 
The constraints on $ h_2 $ are shown in the $ (m_{h_2} \text{ (GeV)},\log_{10}\sin\theta) $ plane. We consider four benchmark points: $ (\alpha_D, \Delta_{\chi}/M_{\chi_1}) =$ 
$(10^{-4}, 0.1)$ (upper-left), $(10^{-1}, 0.1)$ (upper-right), $(10^{-4}, 0.2)$ (lower-left), and $(10^{-1}, 0.2)$ (lower-right). In these benchmarks, the mass spectrum is fixed at $ m_{h_2} = \frac{5}{6} m_{A'} = 2.5 M_{\chi_1} $. 
}
\label{fig:case1}
\end{figure}

In this section, we summarize the relevant constraints for the light $h_2$ with the mass spectra $m_{h_2}\sim m_{A^{\prime}} > 2M_{\chi_2}$ which is the main focus of this work: 
\begin{itemize}
\item The mixing angle $\theta$ between $h$ and $h_D$ is constrained by Higgs boson signal strength measurements: $\sin^2\theta\lesssim 0.12$ at $95\%$ confidence level (C.L.)~\cite{Aad:2015pla,Khachatryan:2016vau} (blue region in Fig.~\ref{fig:case1});
 
\item The $h_1$ decay width is measured to be $\Gamma_{h_1}=3.0^{+2.0}_{-1.5}$ MeV~\cite{CMS:2024eka}. Here we conservatively require $ \Gamma_{h_1}\lesssim 7.0 $ MeV (within $ 2\sigma $) as shown in the pink region in Fig.~\ref{fig:case1}; 

\item For the invisible decays of $ h_1 $, an upper limit of 
$BR(h_1\rightarrow inv.) < 0.107$ at $95\% $ C.L. has been set~\cite{ATLAS:2023tkt} (black region in Fig.~\ref{fig:case1}). In our scenarios, "$ inv. $" conservatively includes only the $\chi_1\chi_1$ and $ h_2 h_2\rightarrow (\chi_1\chi_1)(\chi_1\chi_1) $ modes; 

\item Constraints from $B\rightarrow K h_2\rightarrow K (\chi_1\chi_1)$ and $B\rightarrow K^{\ast} h_2\rightarrow K^{\ast} (\chi_1\chi_1)$ are imposed using the upper limits $BR(B^{\pm}\to K^{\pm}\nu\bar{\nu}) < 4.1\times 10^{-5}$ and $BR(B^{\pm}\to K^{\ast\pm}\nu\bar{\nu}) < 7.9\times 10^{-5}$ at $90\%$ C.L. from Belle II and BaBar experiments, respectively~\cite{Belle-II:2021rof,Lees:2013kla} (green and orange regions in Fig.~\ref{fig:case1})\footnote{We do not include the constraints from recent $B^{\pm}\to K^{\pm}\nu\bar{\nu}$ searches from Belle II due to a mild excess~\cite{Belle-II:2023esi}. We plan to examine whether this anomaly can be explained within the framework of inelastic DM models in Sec.~\ref{sec5}.}.  
The formulas for the branching ratios of $ B\rightarrow K h_2 $ and $ B\rightarrow K^{\ast} h_2 $ are provided in Appendix~\ref{appendix:A}; 
\item Constraints from $\Upsilon (1S)\rightarrow h_2\gamma\rightarrow (\chi_1\chi_1)\gamma$ are derived from Belle and BaBar experiments at $90\%$ C.L.~\cite{delAmoSanchez:2010ac,Seong:2018gut} and are indicated by the yellow (Belle) and the orange (BaBar) dashed regions in Fig.~\ref{fig:case1}. The corresponding formula is given in Appendix~\ref{appendix:B};

\item The constraint from $ J/\psi\rightarrow h_2\gamma\rightarrow (\chi_1\chi_1)\gamma $ measured by the BESIII experiment at $90\%$ C.L.~\cite{Ablikim:2020qtn} is shown as the purple dashed region in Fig.~\ref{fig:case1}. The partial decay width for $ J/\psi\rightarrow h_2\gamma $ can be found in Eq.~(\ref{1Sh2a}), with the substitutions $m_b\rightarrow m_c$ and $ M_{\Upsilon (1S)}\rightarrow M_{J/\psi}$;

\item The search for invisible decays of $h_2$ at LEP has been published in Ref.~\cite{Abbiendi:2007ac} and is represented by the red dashed region in Fig.~\ref{fig:case1}.
\end{itemize} 

We consider four benchmark points (BPs), $ (\alpha_D, \Delta_{\chi}/M_{\chi_1}) =$ 
$(10^{-4}, 0.1)$, $(10^{-1}, 0.1)$, $(10^{-4}, 0.2)$, $(10^{-1}, 0.2)$
 for the mass spectrum $m_{h_2} = \frac{5}{6} m_{A'} = 2.5 M_{\chi_1}$ in Fig.~\ref{fig:case1}. These are typical parameter settings for collider studies~\cite{Tucker-Smith:2001myb}. Similarly, we adopt the same four BPs for the mass spectrum $m_{h_2} = 2.5 M_{\chi_1}$ and $m_{A'} = 2.01 M_{\chi_1} + \Delta_{\chi}$ in Fig.~\ref{fig:case2} of Appendix~\ref{appendix:C}. In contrast, these represent specific parameter settings relevant to the resonance annihilation scenario.

It is evident that as $\alpha_D$ increases, the constraints on $\sin\theta$ become more stringent, particularly due to constraints from Higgs boson invisible decays and precision measurements of the Higgs boson width. This behavior can be understood as follows: these couplings $f$, $\lambda_{h_1 h_2 h_2}$ are related to $v_D$. Once the mass spectrum ($m_{h_2}$, $m_{A'}$, and $\Delta_{\chi}$) is fixed, these couplings are directly connected to $\alpha_D$. Consequently, a larger $\alpha_D$ enhances the branching ratios of $h_1\rightarrow\chi_{1,2}\chi_{1,2}$ and $h_1\rightarrow h_2 h_2$, leading to more stringent constraints on $\sin\theta$.

Because the internal parameters are closely interconnected once the mass spectrum is fixed, it is not feasible to choose large values for both $\alpha_D$ and $\sin\theta$ simultaneously for phenomenological studies while satisfying the current constraints on $h_2$. Therefore, in the next section, we plan to focus on two search strategies for semivisible decays of the dark Higgs boson at Belle II: one with a sizable $\alpha_D$ and a small $\sin\theta$, and another with a small $\alpha_D$ and moderate $\sin\theta$.

\section{Searching for semivisible dark Higgs boson at Belle II}
\label{sec4}

\subsection{Signal signatures, benchmark points and kinematic distributions}

\begin{figure*}[ht!]
\centering{\includegraphics[width=0.4\textwidth]{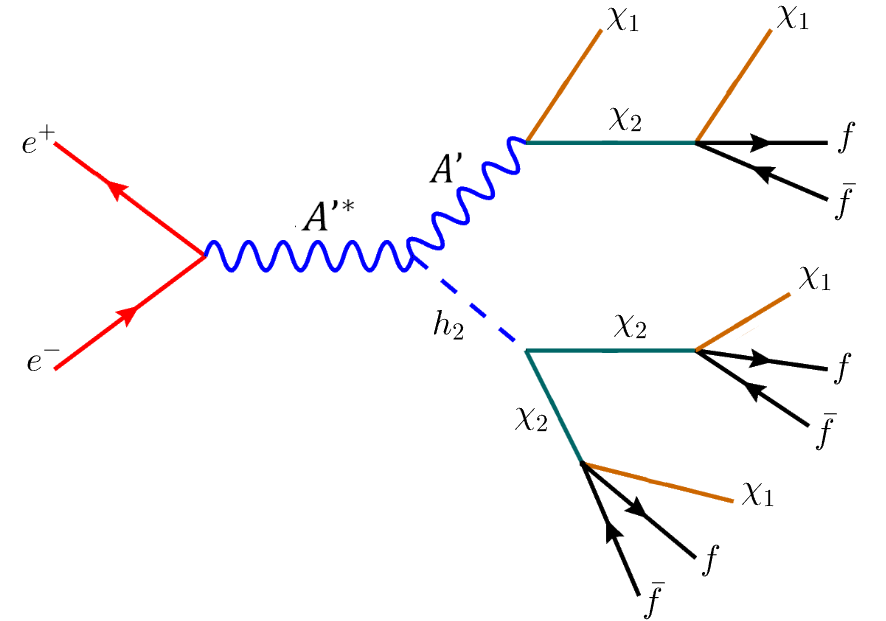}}
\centering{\includegraphics[width=0.50\textwidth]{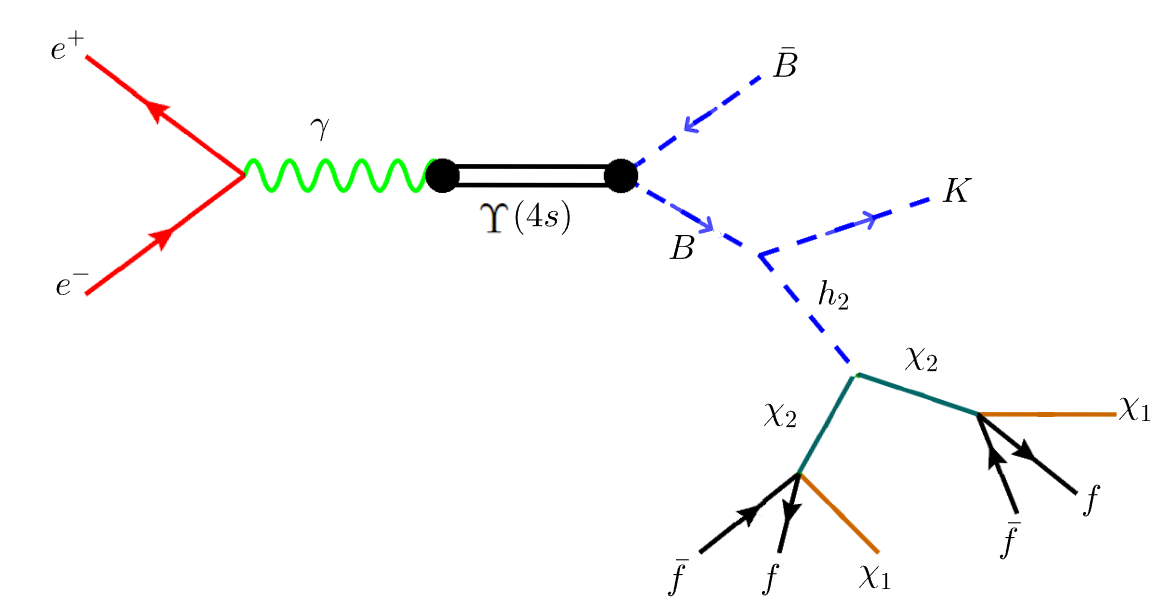}}
\caption{The Feynman diagrams for $e^{+}e^{-}\rightarrow A^{\prime} h_2$, $h_2 \rightarrow \chi_2 \chi_2$, $A'\rightarrow \chi_1\chi_2$ and $e^{+}e^{-}\rightarrow \Upsilon(4s)\rightarrow B \Bar{B}$, $B \rightarrow K h_2$, $h_2 \rightarrow \chi_2 \chi_2$ where $\chi_2\rightarrow\chi_1 f\bar{f}$ for both of these two processes.}
\label{fig:Feyn_diagram}
\end{figure*}

In this study, we focus on two signal processes that produce the dark Higgs boson. The first is the dark Higgs-strahlung process, $e^{+}e^{-}\rightarrow A^{\prime} h_2$, where $h_2\to\chi_2\chi_2$ and $A'\to\chi_1\chi_2$ as depicted in the left panel of Fig.~\ref{fig:Feyn_diagram}. The second is a rare $B$ meson decay, $B \rightarrow K h_2$, with $h_2\to\chi_2\chi_2$, as shown in the right panel of Fig.~\ref{fig:Feyn_diagram}. The long-lived $\chi_2$ subsequently decays into $\chi_1$ and a pair of SM fermions. Consequently, we refer to this distinct signal signature as semivisible decays of the dark Higgs boson. For the dark Higgs-strahlung process, the production cross-section is sensitive to both $\alpha_D$ and $\epsilon$ but relatively insensitive to $\sin\theta$. In contrast, the production cross-section for the rare $B$-meson decay process is primarily sensitive to $\sin\theta$ and independent of $\alpha_D$ and $\epsilon$. Therefore, studying these two signal processes enables us to probe complementary regions of the parameter space. 
  
\begin{figure*}[ht!]
\centering{\includegraphics[width=1\textwidth]{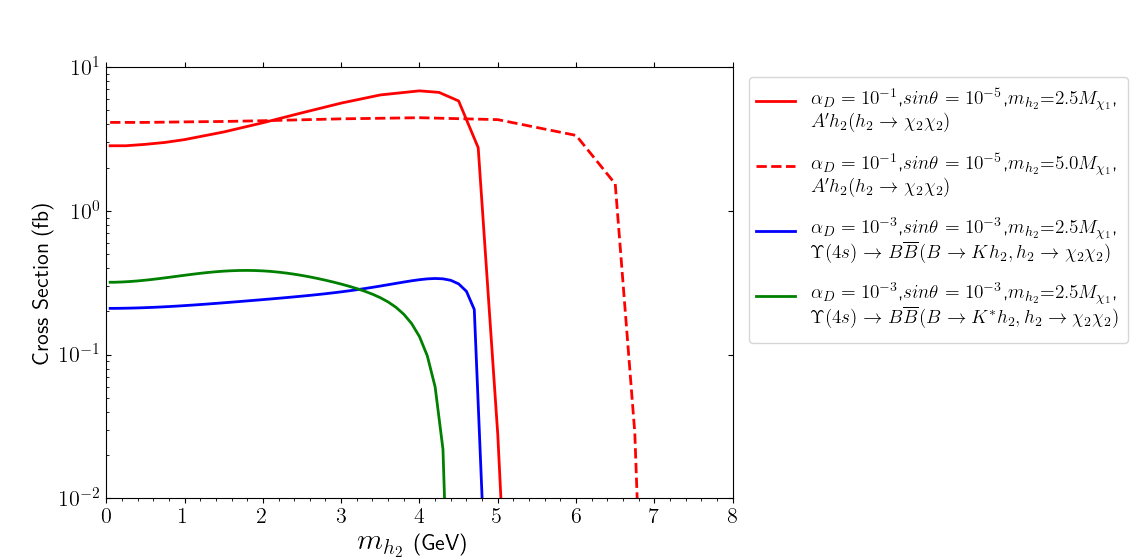}}
\caption{The relations between $m_{h_2}$ (GeV) and $\sigma$ (fb) for $e^{+}e^{-}\rightarrow A^{\prime} h_2, h_2 \rightarrow \chi_2 \chi_2$ (red lines) and $e^{+}e^{-}\rightarrow \Upsilon(4s)\rightarrow B \Bar{B}, B \rightarrow K h_2, h_2 \rightarrow \chi_2 \chi_2$ (blue line) and $e^{+}e^{-}\rightarrow \Upsilon(4s)\rightarrow B \Bar{B}, B \rightarrow K^{\ast} h_2, h_2 \rightarrow \chi_2 \chi_2$ (green line). Here, the other free parameters are fixed: $m_{A^{\prime}} = 3 M_{\chi_1}$, $\alpha_D = g^{2}_D/4\pi = 0.1$, $\epsilon = 10^{-3}$, and $\Delta_{\chi}=0.1M_{\chi_1}$ for these signal processes.}
\label{fig:DM_cross_section}
\end{figure*}

To estimate the production cross-sections for the two processes depicted in Fig.~\ref{fig:Feyn_diagram} at Belle II (with $E(e^+)=4.0$ GeV and $E(e^-)=7.0$ GeV), we first generate the UFO model file using FeynRules~\cite{Ducati:2013cga}. We then calculate the cross-sections for the process $e^{+}e^{-}\rightarrow A^{\prime} h_2$ (with $h_2 \rightarrow \chi_2 \chi_2$) using \textbf{MadGraph5\_aMC@NLO}~\cite{Alwall:2011uj}, setting $\alpha_D=0.1$, $\sin\theta=10^{-5}$, and choosing either $m_{h_2}=2.5M_{\chi_1}$ or $m_{h_2}=5.0M_{\chi_1}$. The results are displayed in Fig.~\ref{fig:DM_cross_section} as a solid red line for $m_{h_2}=2.5M_{\chi_1}$ and a dashed red line for $m_{h_2}=5.0M_{\chi_1}$. %
However, due to non-perturbative effects in the process $e^{+}e^{-}\rightarrow \Upsilon(4S) \rightarrow B \Bar{B}$, we calculate its cross-section as follows. First, at the Belle II, the production cross-section for $\Upsilon(4s)$ is approximately $1.1$ nb~\cite{Nishimura:2012ep}. Next, the on-shell $\Upsilon(4s)$ decays into a pair of $B$ mesons with a $96\%$ probability~\cite{CLEO:2000sdf}. We then compute the branching ratios of $B\to K^{(\ast)}h_2$ using the formula provided in the Appendix~\ref{appendix:A}. Finally, the branching ratio for $h_2\to\chi_2\chi_2$ is determined using Eq.~(\ref{eq:h2 decay eq.}). In these calculations, the parameters $\alpha_D=10^{-3}$, $\sin\theta=10^{-3}$, and $m_{h_2}=2.5M_{\chi_1}$ are fixed. The outcomes are shown as the blue solid line for $B \rightarrow K h_2$ and the green solid line for $B \rightarrow K^{\ast} h_2$ in Fig.~\ref{fig:DM_cross_section}. The other free parameters are fixed as follows: $m_{A^{\prime}} = 3 M_{\chi_1}$, $\alpha_D = g^{2}_D/4\pi = 0.1$, $\epsilon = 10^{-3}$, and $\Delta_{\chi}=0.1M_{\chi_1}$ for these signal processes. We find that the cross-sections for the dark Higgs-strahlung processes are generally larger than those for the rare $B$ meson decay processes, although they probe different regions of parameter space, as discussed above.

\begin{figure*}[ht!]
\centering{\includegraphics[width=1.0\textwidth]{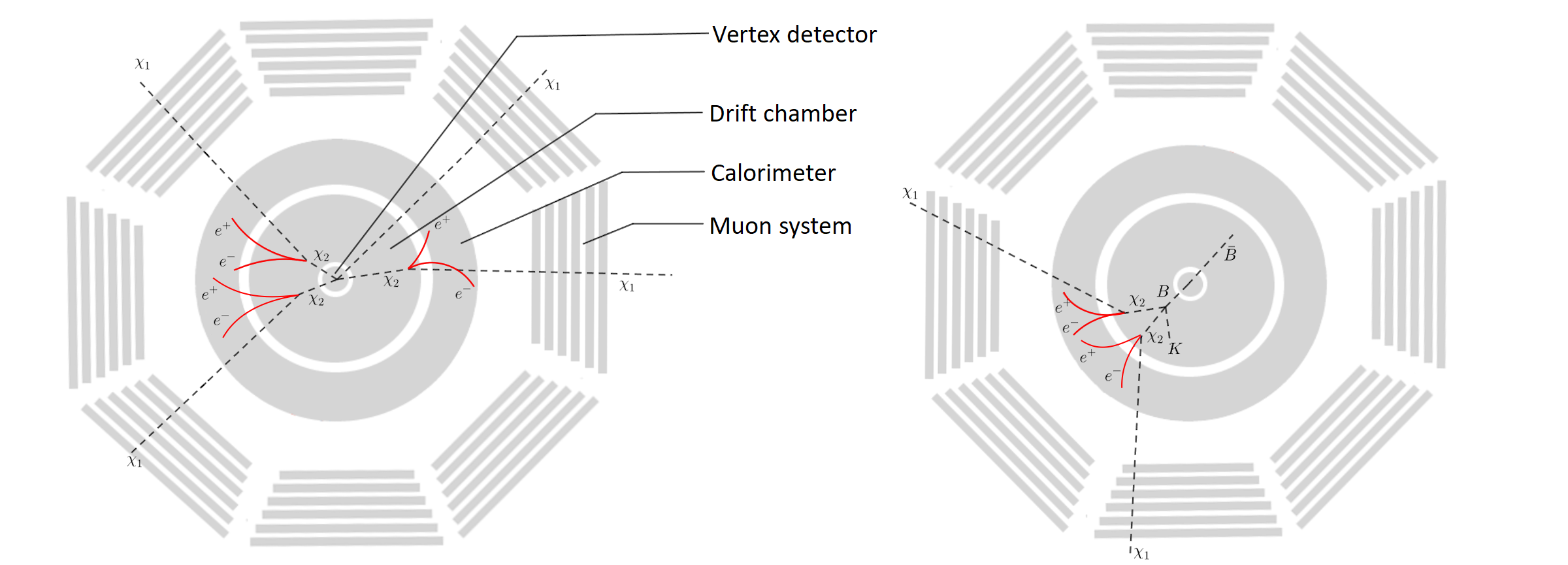}}
\caption{The schematic diagrams for the signature of the dark Higgs-strahlung process (left panel) and the signature of a rare $B$ meson decay process (right panel) in inelastic DM models at Belle II.}
\label{fig:Detector_diagram}
\end{figure*}

The structure of the Belle II detector is presented in Fig.~\ref{fig:Detector_diagram} designed for accurate particle tracking, momentum measurement, and identification. It comprises several sub-detectors positioned concentrically around the interaction point (IP), including the Vertex Detector, Drift Chamber, Calorimeter, and Muon System~\cite{Adachi:2018qme}. 
In this study, we investigate the detection of LLPs in inelastic DM models at Belle II. Specifically, the left panel of Fig.~\ref{fig:Detector_diagram} illustrates the signature of the dark Higgs-strahlung process, while the right panel displays the signature of a rare $B$ meson decay process.

\begin{figure*}[ht!]
\centering{\includegraphics[width=0.42\textwidth]{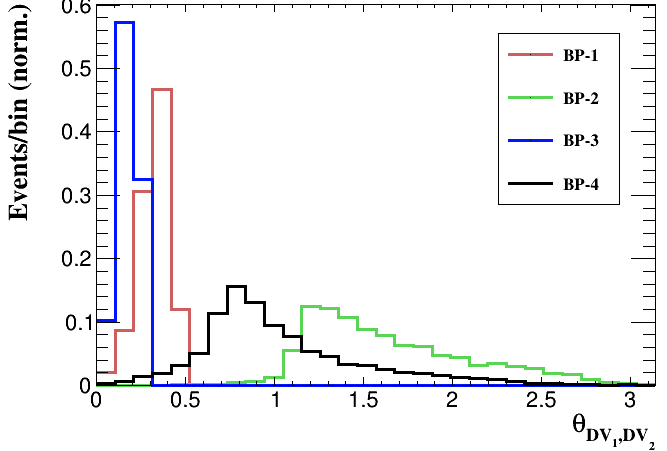}}
\centering{\includegraphics[width=0.42\textwidth]{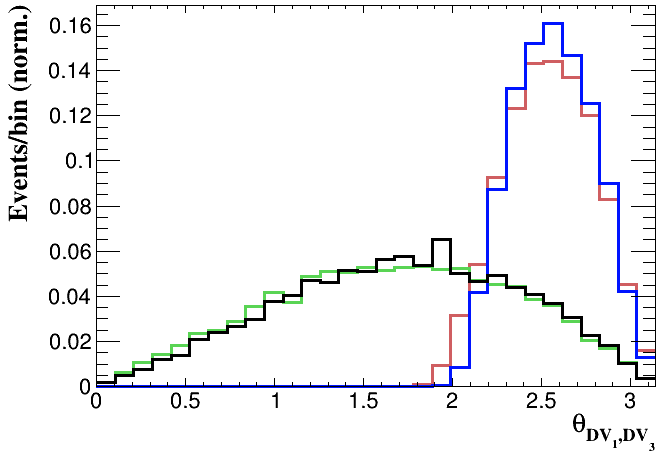}}
\centering{\includegraphics[width=0.42\textwidth]{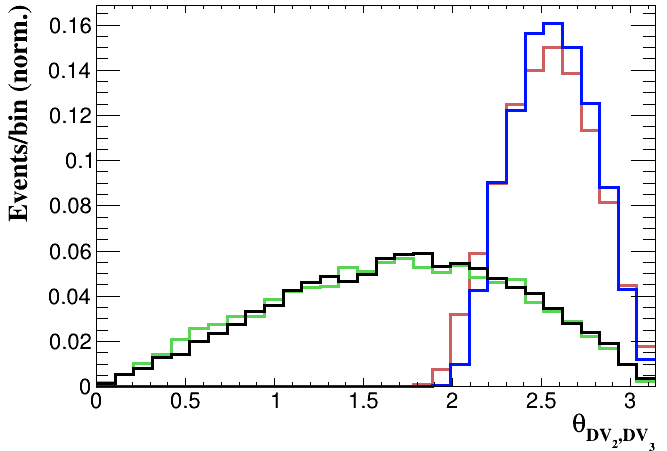}}
\centering{\includegraphics[width=0.42\textwidth]{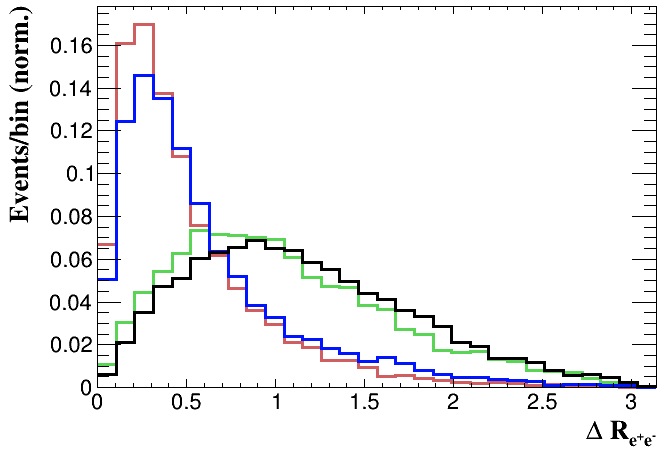}}
\centering{\includegraphics[width=0.42\textwidth]{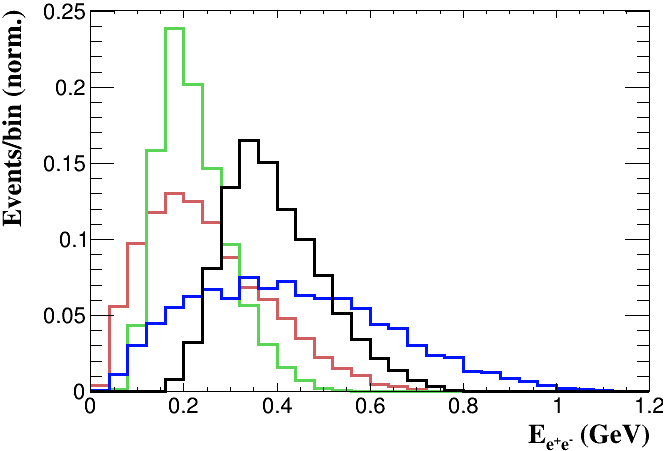}}
\centering{\includegraphics[width=0.42\textwidth]{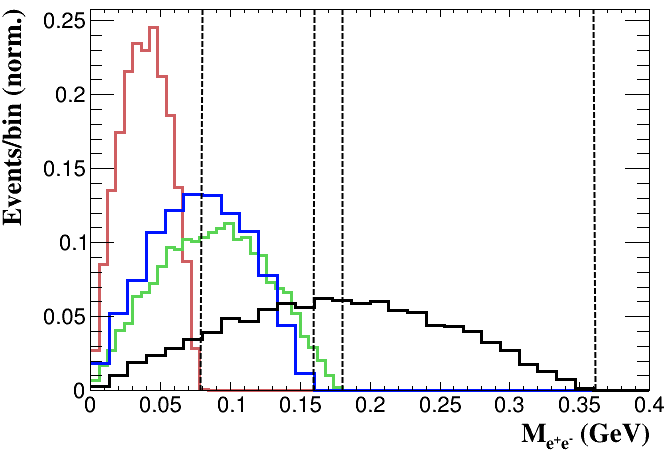}}
\centering{\includegraphics[width=0.42\textwidth]{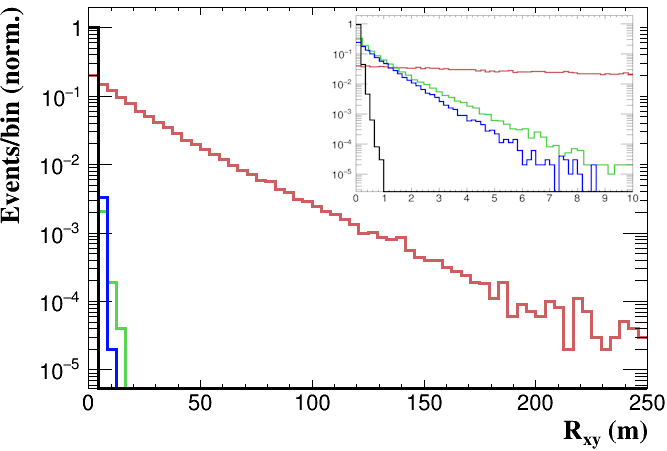}}
\centering{\includegraphics[width=0.42\textwidth]{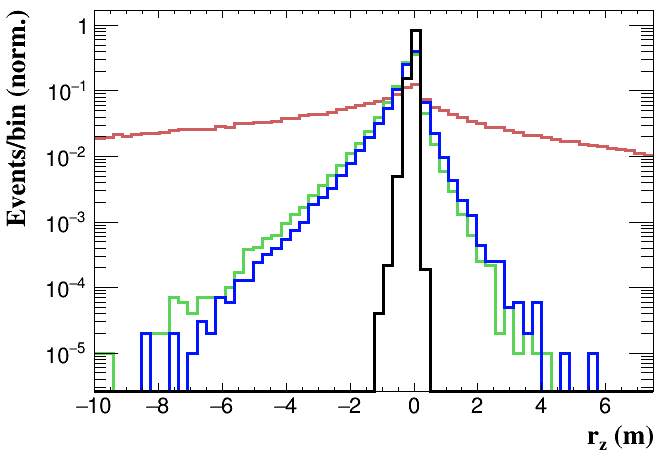}}
\caption{Various kinematic distributions for $e^{+}e^{-} \rightarrow A^{\prime} h_2\rightarrow (\chi_1\chi_2)(\chi_2\chi_2)$ with $\chi_2 \rightarrow \chi_1 f\bar{f}$ based on the BP-1 to BP-4 mentioned in
the main text. In order, they are open angles of the three displaced vertices, $\theta_{DV_{i},DV_{j}}$, angular distance between electron pair, $\Delta R_{e^{+}e^{-}}$, electron pair energy, $E_{e^{+}e^{-}}$ (GeV), electron pair invariant mass, $M_{e^{+}e^{-}}$ (GeV), the transverse distribution of displaced vertices, $R_{xy}$ (m), and the longitudinal distribution of displaced vertices, $r_{z}$ (m).}
\label{fig:DM_kineDis_1}
\end{figure*}

For the dark Higgs-strahlung process, we use the following four BPs with $m_{h_2} = \frac{5}{6}m_{A^{\prime}} = 2.5 M_{\chi_1}$ to display our analysis:
\begin{itemize}
    \item BP-1: $m_{h_2} = 2.0 $ GeV, $\Delta_{\chi} = 0.1 M_{\chi_1}$, $\epsilon = 2\times10^{-3}$; 
    \item BP-2: $m_{h_2} = 4.5 $ GeV, $\Delta_{\chi} = 0.1 M_{\chi_1}$, $\epsilon = 2\times10^{-3}$; 
    \item BP-3: $m_{h_2} = 2.0 $ GeV, $\Delta_{\chi} = 0.2 M_{\chi_1}$, $\epsilon = 10^{-3}$; 
    \item BP-4: $m_{h_2} = 4.5 $ GeV, $\Delta_{\chi} = 0.2 M_{\chi_1}$, $\epsilon = 10^{-3}$.
\end{itemize} 
With a fixed $\alpha_D = 0.1$, we present typical kinematic distributions in Fig.~\ref{fig:DM_kineDis_1}. 

Firstly, in the dark Higgs-strahlung process, three displaced vertices (DVs) are produced. By comparing the opening angles between each pair of these vertices, we observe that the distributions of $\theta_{DV_{1},DV_{3}}$ and $\theta_{DV_{2},DV_{3}}$ are nearly identical, whereas the distribution of $\theta_{DV_{1},DV_{2}}$ differs significantly. This indicates that $DV_1$ and $DV_2$ originate from the decay of $h_2$ into a pair of $\chi_2$, while $DV_3$ is associated with the decay of the dark photon into $\chi_1\chi_2$. A smaller $m_{h_2}$ leads to larger values of $\theta_{DV_{1},DV_{3}}$ and $\theta_{DV_{2},DV_{3}}$ because a lower $m_{h_2}$ makes it easier for both $A^{\prime}$ and $h_2$ to be boosted. This increased boost causes the opening angle $\theta_{DV_{1},DV_{3}}$ and $\theta_{DV_{2},DV_{3}}$ to approach $180$ degrees. Conversely, for the $\theta_{DV_{1},DV_{2}}$ distribution, a smaller $m_{h_2}$ brings the two decay vertices closer together. The mass splitting $\Delta_{\chi}$ only slightly influence the distributions of $\theta_{DV_{1},DV_{3}}$ and $\theta_{DV_{2},DV_{3}}$, but it significantly affects the $\theta_{DV_{1},DV_{2}}$ distribution. This is because a larger $\Delta_{\chi}$ imparts more energy to each $e^+e^-$ pair 
produced from the decays of the two $\chi_2$ particles originating from $h_2$, thereby reducing  $\theta_{DV_{1},DV_{2}}$.

Each of the two displaced vertices resulting from $h_2$ decay produces a pair of opposite-sign electrons. The angular distribution $\Delta R_{e^{+}e^{-}}$ primarily depends on $m_{h_2}$, with only a mild influence from $\Delta_{\chi}$. When $m_{h_2}$ is held constant, a smaller $\Delta_{\chi}$ causes a slightly leftward shift in the $\Delta R_{e^{+}e^{-}}$ distribution. Furthermore, the energy distribution of the electron pairs indicates that, for a given $\Delta_{\chi}$, a smaller $m_{h_2}$ leads to broader energy distributions. Additionally, the invariant mass distribution $M_{e^{+}e^{-}}$ exhibits a threshold at $\Delta_{\chi}$. When both $m_{h_2}$ and $\Delta_{\chi}$ are small, the displaced vertices are located further from the IP. These observations are also evident in the distributions of $R_{xy}$ and $r_z$.

As shown in Fig.~\ref{fig:Kine_theta} of Appendix~\ref{appendix:C}, when $m_{h_2}$ exceeds approximately $4$ GeV, the overlap between $\theta_{DV_{1},DV_{2}}$ and $\theta_{DV_{1},DV_{3}}$ can reach up to $30\%$. This indicates that the $\chi_2$ vertex produced by $A^{\prime}$ decays and those from $h_2$ decay are difficult to distinguish. This behavior arises because both $m_{h_2}$ and $m_{A'}$ increase simultaneously under our parameter settings, reducing the likelihood that the two displaced vertices will be boosted back-to-back in the longitudinal direction. Consequently, the distributions of $\theta_{DV_1,DV_3}$ and $\theta_{DV_2,DV_3}$ shift slightly to the left. Moreover, since $h_2$ is heavier and less easily accelerated, its decay yields a broader distribution for $\theta_{DV_1,DV_2}$. Therefore, when $m_{h_2}$ exceeds about $4$ GeV, it cannot be avoided to account for the possibility of $\chi_2$ vertex originating from $A^{\prime}$ decay in our event selections.

For the case where $m_{A'} < M_{\chi_1} + M_{\chi_2}$, the dark photon $A^{\prime}$ decays into a pair of SM fermions. Moreover, $A^{\prime}$ can behave as a LLP if the kinetic mixing parameter $\epsilon$ is sufficiently small. Since the invariant mass and energy of this displaced vertex differ from those of the displaced vertex produced by $\chi_2$, and no additional missing energy is present. Consequently, this displaced vertex can be effectively identified and distinguished from those from $\chi_2$ decay. This situation is similar to the study presented in Ref.~\cite{Duerr:2020muu}, so we do not focus on the details of this analysis in the present work. 

For the rare $B$ meson decay process, we examine four additional BPs with $m_{h_2} = 2.5 M_{\chi_1}$ and $m_{A^{\prime}} = 2.01 M_{\chi_1} + \Delta_{\chi}$ (close-to-resonance scenario) to illustrate our analysis:
\begin{itemize}
    \item BP-A: $m_{h_2} = 2.0 $ GeV, $\Delta_{\chi} = 0.1 M_{\chi_1}$, $\epsilon = 2\times10^{-2}$; 
    \item BP-B: $m_{h_2} = 4.5 $ GeV, $\Delta_{\chi} = 0.1 M_{\chi_1}$, $\epsilon = 2\times10^{-2}$; 
    \item BP-C: $m_{h_2} = 2.0 $ GeV, $\Delta_{\chi} = 0.2 M_{\chi_1}$, $\epsilon = 10^{-2}$; 
    \item BP-D: $m_{h_2} = 4.5 $ GeV, $\Delta_{\chi} = 0.2 M_{\chi_1}$, $\epsilon = 10^{-2}$.
\end{itemize}
Here, $\alpha_D = 10^{-3}$ is fixed. 
The cross-section is independent of $\epsilon$ in the rare $B$ meson decay process. However, the lifetime of $\chi_2$ is inversely proportional to $\epsilon^{2}$. Unlike the dark Higgs-strahlung process, we set $\epsilon = 10^{-2}$ to $2\times 10^{-2}$ in the rare $B$ meson decay analysis. This ensures that $\chi_2$ does not travel too far, allowing a sufficient number of signal events ($N_{\text{event}}$) to be detected within the detector.

\begin{figure*}[ht!]
\centering{\includegraphics[width=0.48\textwidth]{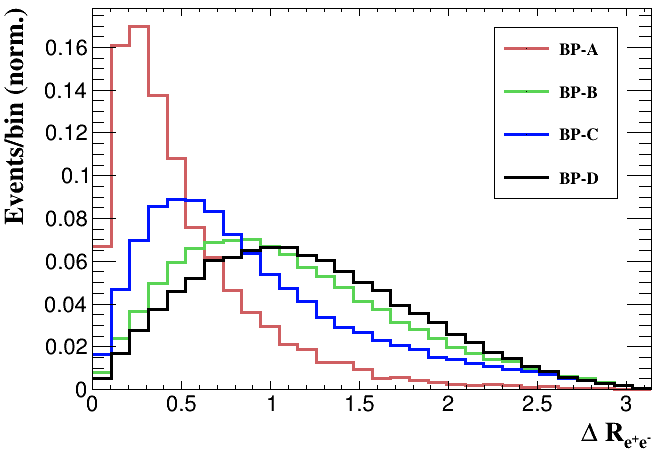}}
\centering{\includegraphics[width=0.48\textwidth]{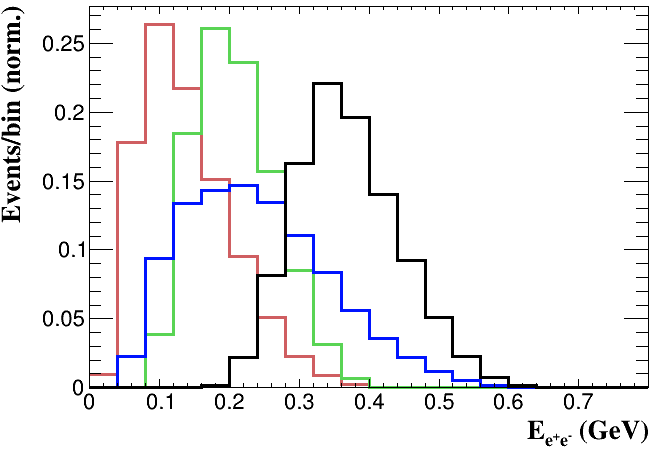}}
\centering{\includegraphics[width=0.48\textwidth]{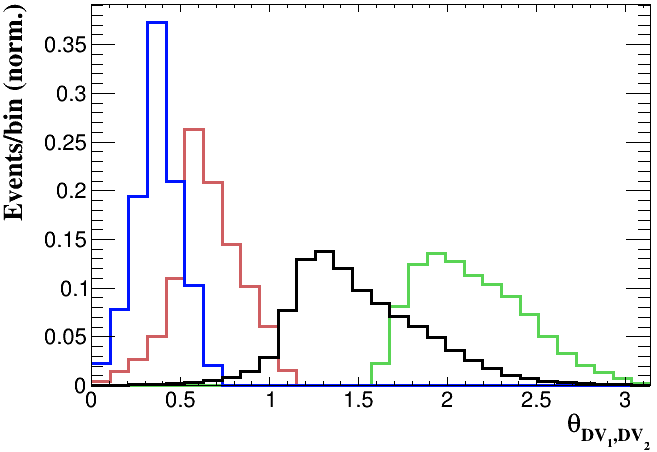}}
\centering{\includegraphics[width=0.48\textwidth]{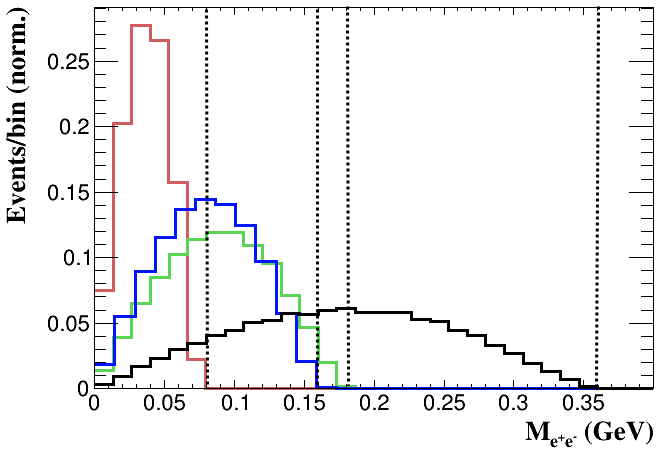}}
\centering{\includegraphics[width=0.48\textwidth]{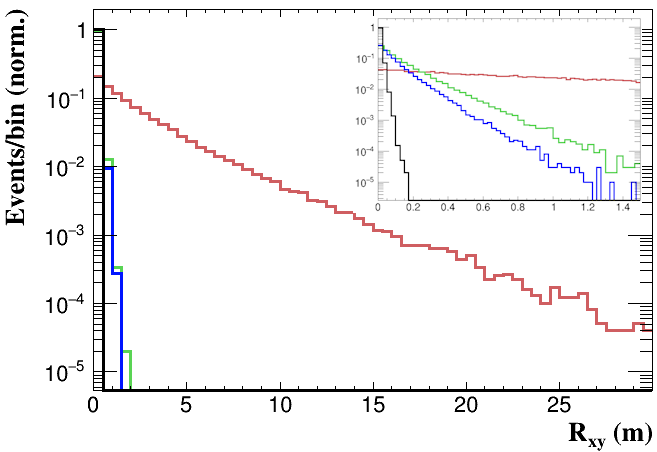}}
\centering{\includegraphics[width=0.48\textwidth]{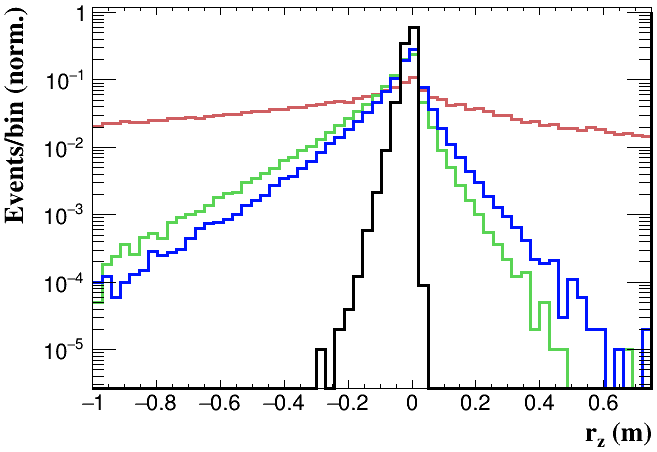}}
\caption{ Various kinematic distributions for $e^{+}e^{-}\rightarrow \Upsilon(4s)\rightarrow B \Bar{B} (B \rightarrow K h_2, h_2 \rightarrow \chi_2 \chi_2)$ with $\chi_2 \rightarrow \chi_1 f\bar{f}$ based on the BP-A to BP-D mentioned in
the main text. In order, they are angular distance between electron pair, $\Delta R_{e^{+}e^{-}}$, electron pair energy, $E_{e^{+}e^{-}}$ (GeV), open angles of the two displaced vertices, $\theta_{DV_{1},DV_{2}}$, electron pair invariant mass,  $M_{e^{+}e^{-}}$ (GeV), the tranverse distribution of displaced vertices, $R_{xy}$ (m), and the longitudinal distribution of displaced vertices, $r_{z}$ (m).}
\label{fig:DM_kineDis_3}
\end{figure*}

Some key kinematic distributions are shown in Fig.~\ref{fig:DM_kineDis_3}. 
Firstly, Unlike the dark Higgs-strahlung process, in the rare $B$ meson decay process, the $\Delta R_{e^{+}e^{-}}$ distribution of the electron pair from $\chi_2$ decay is more influenced by the mass splitting $\Delta_\chi$. A larger $\Delta_\chi$ results in an increased $\Delta R_{e^{+}e^{-}}$. This difference arises because, in the dark Higgs-strahlung process, $h_2$ is produced directly from $e^{+}e^{-}$ collisions, whereas in the rare $B$ meson decay process, $h_2$ originates from $\Upsilon(4s)$ decay into $B \bar{B}$, followed by $B$ meson decay. Consequently, $h_2$ has lower energy, leading to a less boosted $\chi_2$ and making the $\Delta R_{e^{+}e^{-}}$ distribution more sensitive to $\Delta_\chi$.
Secondly, from the electron pair energy distribution $E_{e^{+}e^{-}}$, we observe that in the rare $B$ meson decay process, $E_{e^{+}e^{-}}$ is lower when $\Delta_\chi$ and $m_{h_2}$ are fixed. Additionally, due to the lower energy of $h_2$, the angle between the two displaced vertices from $\chi_2$ of $h_2$ decays, $\theta_{DV_1, DV_2}$, in the rare $B$ meson decay process is larger than that in the dark Higgs-strahlung process. The distribution patterns of $M_{e^{+}e^{-}}$, $R_{xy}$ and $r_z$ are the same as those of the dark Higgs-strahlung process.

\subsection{Event selections and results}

\begin{table}[h]
\centering
\begin{tabular}{|c|>{\raggedright\arraybackslash}p{10cm}|}
\hline
Objects & Selections \\ \hline
\multirow{2}{*}{displaced vertex} 
  & (i) $-55$ cm $ \leq z \leq 140$ cm \\ \cline{2-2}
  & (ii) $ 17^{\circ}\leq \theta_{LAB}^{DV} \leq 150^{\circ}$ \\ \hline
  & (i) both $E(e^{+})$ and $E(e^{-}) > 0.06$ GeV\\ \cline{2-2}
\multirow{3}{*}{electrons} & (ii) At least a pair of $e^{+}e^{-}$ with opening angle of pair $\theta_{e^{+}e^{-}} > 0.025$ rad \\ \cline{2-2}
  & (iii) At least a pair of $e^{+}e^{-}$ with invariant mass of pair $M_{e^{+}e^{-}} > 0.03$ GeV \\ \hline
\end{tabular}
\caption{Event selections for the displaced vertex analysis at Belle II.}
\label{tab:cut_flow_1}
\end{table}

We closely follow Refs.~\cite{Duerr:2019dmv,Fuchs:2020cmm} to set up event selections for the displaced vertex signatures at Belle II. 
We assume that all background events can be suppressed to a negligible level considering the above choices and that the $\chi_2$ displaced vertex is detected with 100\% efficiency at the vertex detector and drift chamber of Belle II. Therefore, we consider the following two background-free regions:
\begin{itemize}
    \item Low $R_{xy}$ region: 0.2 $< R_{xy} \leq$ 17.0 cm;
    \item High $R_{xy}$ region: 17.0 $< R_{xy} \leq$ 60.0 cm.
\end{itemize} 
Table~\ref{tab:cut_flow_1} summarizes the event selection criteria applied in our analysis. 
In our study, within the considered $\chi_2$ mass range, $\chi_2$ cannot decay into $\chi_1$ plus a muon pair. Therefore, we only consider the decay process $\chi_2 \rightarrow \chi_1 e^{+} e^{-}$. First, we require that both long-lived $\chi_2$ particles (produced in $h_2$ decays) decay within the detector volume, with at least two displaced vertices detected. To suppress possible background events, all electrons and positrons from both displaced vertices must individually satisfy $E(e^{\pm}) > 0.06$ GeV. Additionally, at least one pair of opposite-sign electrons must have an opening angle greater than $0.025$ rad and an invariant mass $M_{e^{+}e^{-}}$ exceeding $0.03$ GeV.

The results for BP-1 to BP-4 in the dark Higgs-strahlung process are shown in Table~\ref{tab:DM_1}. Similarly, BP-A to BP-D for the rare $B$ meson decay process are presented in Table~\ref{table_DM_cut_flow_2}. 
Here, Eff.(low $R_{xy}$) and Eff.(high $R_{xy}$) represent the efficiencies for the low and high $R_{xy}$ regions after event selections. An integrated luminosity of 1 fb$^{-1}$ is used to compute $N_{\text{event}}$. The results indicate that the efficiencies in the low and high $R_{xy}$ regions are closely related to the spatial distribution of $\chi_2$ displaced vertices and the flight distance of $\chi_2$. 
As $M_{\chi_2}$ and $\Delta_{\chi}$ increase, the flight distance of $\chi_2$ decreases, leading to a higher number of $N_{\text{event}}$. Notably, when transitioning from BP-3 to BP-4 (or from BP-C to BP-D), Eff.(low $R_{xy}$) increases, while Eff.(high $R_{xy}$) decreases.

\begin{table}[h!]
\centering
\begin{tabular}{|c|c|c|c|c|c|}
\hline
BP & $\sigma$ (fb) & Eff.(low $R_{xy}$) & Eff.(high $R_{xy}$) & $N_{\text{event}}$ \\ \hline
  BP-1 & 2.58 & 0.001\% & 0.004\% & 20.8 \\ \hline 
  BP-2 & 3.72 & 0.35\% & 1.23\% & $2.68\times10^{3}$ \\ \hline
  BP-3 & 0.73 & 2.63\% & 6.12\% & $6.24\times10^{3}$ \\ \hline 
  BP-4 & 1.03 & 43.49\% & 0.081\% & $3.14\times10^{4}$ \\  \hline
\end{tabular}
\caption{The analysis results of four BPs for the $e^{+}e^{-} \rightarrow A^{\prime} h_2\rightarrow (\chi_1\chi_2)(\chi_2\chi_2)$ with $\chi_2 \rightarrow \chi_1 f\bar{f}$ process.}
\label{tab:DM_1}
\end{table}

\begin{table}[h!]
\centering
\begin{tabular}{|c|c|c|c|c|c|}
\hline
BP & $\sigma$ (fb) & Eff.(low $R_{xy}$) & Eff.(high $R_{xy}$) & $N_{\text{event}}$ \\ \hline
  BP-A & 0.62 & 0.009\% & 0.035\% & $9.17$ \\ \hline 
  BP-B & 0.31 & 1.99\% & 3.11\% & $5.51\times10^{2}$ \\ \hline 
  BP-C & 0.17 & 2.72\% & 3.84\% & $3.97\times10^{2}$ \\ \hline 
  BP-D & 0.087 & 33.15\% & 0.029\% & $1.02\times10^{3}$ \\  \hline
\end{tabular}
\caption{The analysis results of other four BPs for the $e^{+}e^{-}\rightarrow \Upsilon(4s)\rightarrow B \Bar{B} (B \rightarrow K h_2, h_2 \rightarrow \chi_2 \chi_2)$ with $\chi_2 \rightarrow \chi_1 f\bar{f}$ process.}
\label{table_DM_cut_flow_2}
\end{table}

\begin{figure*}[ht!]
\centering{\includegraphics[width=0.9\textwidth]{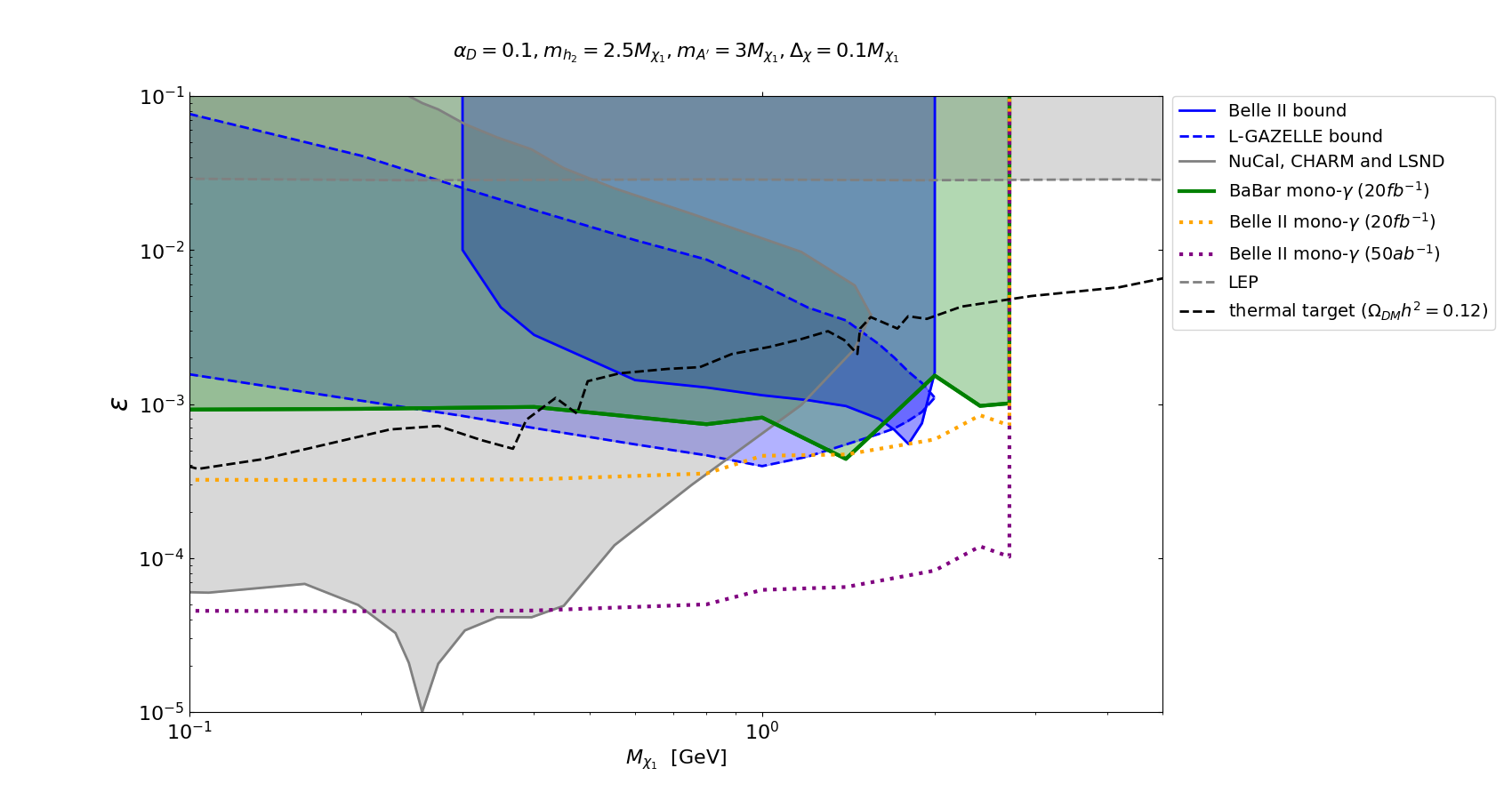}}
\centering{\includegraphics[width=0.9\textwidth]{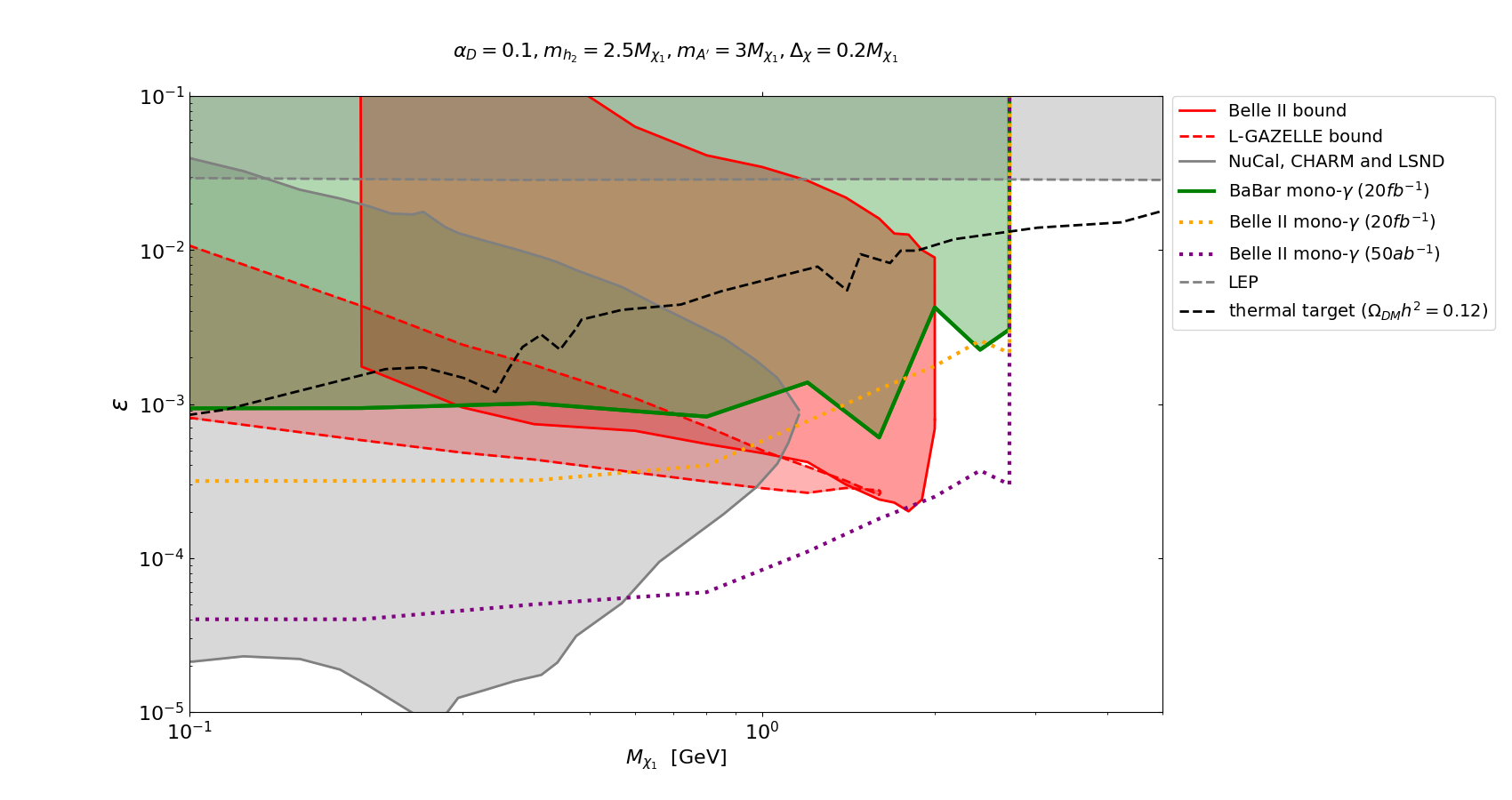}}
\caption{The future bounds from dark Higgs-strahlung process with the integrated luminosity of 50 $ab^{-1}$ (Belle II, solid lines and L-GAZELLE, dashed lines). Here parameters $\alpha_{D} = 0.1$, $m_{h_2} = 2.5 M_{\chi_1}$, $m_{A^{\prime}} = 3M_{\chi_1}$ and $\Delta_{\chi} = 0.1M_{\chi_1}$(blue region), $\Delta_{\chi} = 0.2M_{\chi_1}$(red region) are fixed and $90\%$ C.L. contours which correspond to an upper limit of 2.3 events with the assumption of background-free are applied. The NuCal, CHARM, and LSND bounds~\cite{Tsai:2019buq} are represented by the gray solid line border area. The BaBar mono-$\gamma$ bound~\cite{Frumkin:2022ror} is shown as the green region. The LEP bound~\cite{LinearColliderVision:2025hlt} is depicted by the gray dashed line border area. The projected Belle II mono-$\gamma$ bounds with 20 $fb^{-1}$ (orange dotted lines) and 50 $ab^{-1}$ (purple dotted lines) are included for the comparison. The thermal targets ($\Omega_{DM}h^{2}=0.12$)~\cite{Duerr:2019dmv} are indicated with black dashed lines. }
\label{fig:bounds}
\end{figure*}

Finally, we adopt an optimistic value of 50 ab$^{-1}$ for integrated luminosity, applying the event selections in Table~\ref{tab:cut_flow_1} to fermionic inelastic DM models at Belle II, and predict future bounds for the dark Higgs-strahlung process, shown in Fig.~\ref{fig:bounds} (solid lines, blue region and red region). 
In this analysis, we fix parameters $\alpha_D = 0.1$, $m_{h_2} = 2.5M_{\chi_1}$, $m_{A^{\prime}} = 3M_{\chi_1}$, and consider the mass splittings $\Delta_{\chi} = 0.1M_{\chi_1}$ (blue region) and $\Delta_{\chi} = 0.2M_{\chi_1}$ (red region) with $90\%$ confidence level (C.L.) contours, corresponding to an upper limit of 2.3 events under a background-free assumption. Notice constraints from model-independent NuCal, CHARM and LSND bound~\cite{Tsai:2019buq}, BaBar mono$-\gamma$ bound~\cite{Frumkin:2022ror}, LEP bound~\cite{LinearColliderVision:2025hlt} and the predicted thermal relic abundance ($\Omega_{DM}h^{2}=0.12$)~\cite{Duerr:2019dmv} are added for the comparison.

\begin{figure*}[ht!]
\centering{\includegraphics[width=0.9\textwidth]{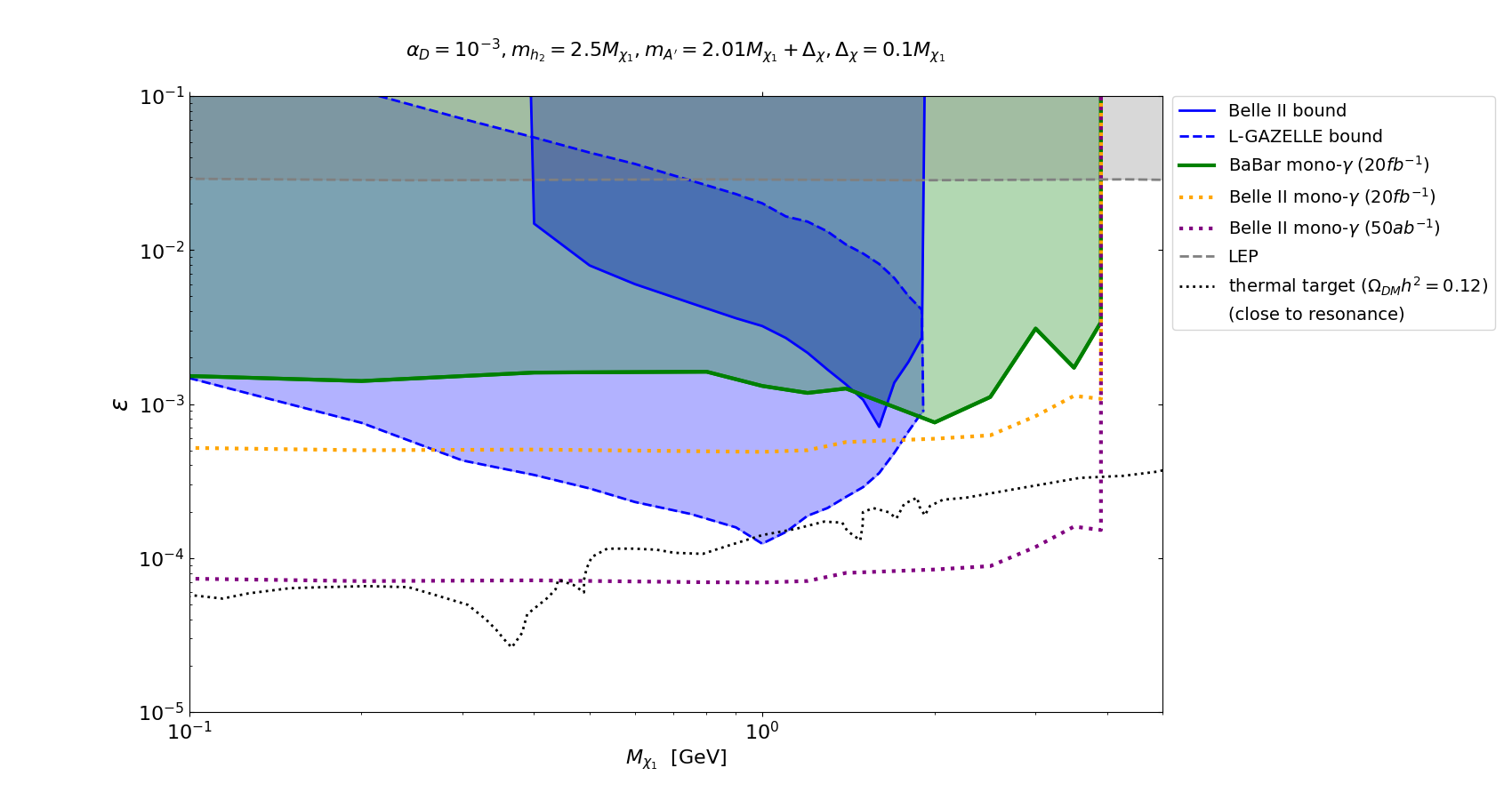}}
\centering{\includegraphics[width=0.9\textwidth]{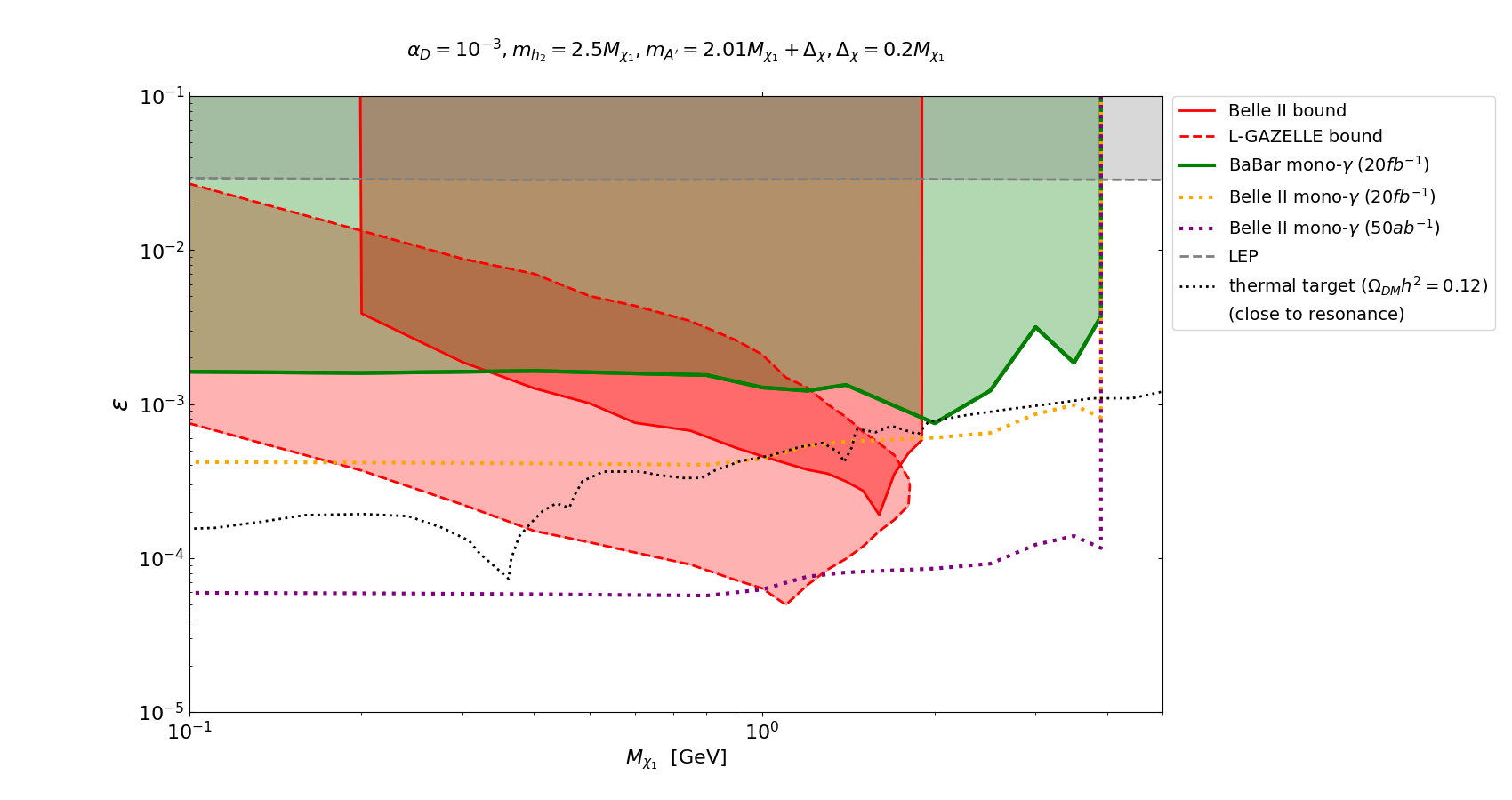}}
\caption{The future bounds from rare $B$ meson decay process with the integrated luminosity of 50 $ab^{-1}$(Belle II, solid lines and L-GAZELLE, dashed lines). Here parameters $\alpha_{D} = 10^{-3}$, $m_{h_2} = 2.5 M_{\chi_1}$, $m_{A^{\prime}} = 2.01M_{\chi_1}+\Delta_{\chi}$, $\sin\theta=10^{-2}$ and $\Delta_{\chi} = 0.1M_{\chi_1}$(blue region), $\Delta_{\chi} = 0.2M_{\chi_1}$(red region) are fixed and $90\%$ C.L. contours which correspond to an upper limit of 2.3 events with the assumption of background-free are applied. The BaBar mono$-\gamma$ bound~\cite{Frumkin:2022ror} is shown as the green region. The LEP bound~\cite{LinearColliderVision:2025hlt} is depicted by the gray dashed line border area. The projected Belle II mono-$\gamma$ bounds with 20 $fb^{-1}$ (orange dotted lines) and 50 $ab^{-1}$ (purple dotted lines) are included for the comparison. The thermal targets ($\Omega_{DM}h^{2}=0.12$) from the close-to-resonance annihilation~\cite{Duerr:2019dmv} are represented by black dotted lines. }
\label{fig:bounds_2}
\end{figure*}

Furthermore, as expected, the projected mono-photon limits from Belle II can provide important complementary constraints compared to those from displaced vertex searches. We have verified that, within our interested parameter space, the contribution to the mono-photon signature from the process $e^+e^- \to \gamma h_2 A'$, followed by $h_2 \to \chi_1\chi_1$ and $A' \to \chi_1\chi_2$, is much smaller than that from the process $e^+e^- \to \gamma \chi_1 \chi_2$. We therefore focus exclusively on the latter. Following the mono-photon search strategy described in Sec.~3.3 of Ref.~\cite{Duerr:2019dmv}, we apply this approach to our parameter space as shown in Figs.~\ref{fig:bounds} and~\ref{fig:bounds_2}, respectively. Both current (20 $fb^{-1}$; orange dotted) and projected (50 $ab^{-1}$; purple dotted) Belle II integrated luminosities are included for the comparison.

As shown in Fig.~\ref{fig:bounds}, the constrained region for $\Delta_\chi = 0.1 M_{\chi_1}$ begins at $M_{\chi_1} > 0.3$ GeV, while for $\Delta_\chi = 0.2 M_{\chi_1}$, it starts at $M_{\chi_1} > 0.2$ GeV. There are two factors constraining the left boundary of the parameter space. The first arises from the requirement that $M_{e^{+}e^{-}} > 0.03$ GeV. The second occurs when $M_{\chi_2}$ is small, necessitating an increase in $\epsilon$ to ensure a sufficient $N_{\text{event}}$ for the $\chi_2$ decay inside the vertex detector.  
Similarly, when $\Delta_\chi$ is fixed and only $M_{\chi_2}$ increases, its flight distance shortens. Consequently, $\epsilon$ can be gradually reduced, allowing the excluded region of the parameter space to extend downward as $M_{\chi_1}$ increases. This trend continues until $M_{\chi_1} \approx 1.8$ GeV, where the parameter space covers the region with the smallest $\epsilon$. Beyond this point, as $M_{\chi_1}$ further increases, the cross-section of this process decreases sharply, as illustrated in Fig.~\ref{fig:DM_cross_section}. To maintain a sufficient $N_{\text{event}}$, $\epsilon$ must increase rapidly, reaching its maximum at $M_{\chi_1} = 2$ GeV, which defines the right boundary of the constrained parameter space. It should be noted that for the dark Higgs-strahlung process, the cross-section is proportional to $\epsilon^2$. Compared to $\Delta_\chi = 0.1M_{\chi_1}$, when $\Delta_\chi = 0.2M_{\chi_1}$, $\chi_2$ has a larger decay width and a shorter flight distance. As a result, for the same $M_{\chi_1}$, the lower limit of the excluded region in the parameter space corresponds to a smaller $\epsilon$. Meanwhile, this also constrains the upper limit to $\epsilon < 0.1$, forming the upper boundary of the excluded region along the red solid line in Fig.~\ref{fig:bounds}.

As shown in Fig.~\ref{fig:bounds}, the exclusion regions accessible to the Belle II detector extend significantly beyond the constraints set by NuCal, CHARM, and LSND, the LEP bounds for both mass splitting scenarios considered, particularly in the region $0.5~\text{GeV} < M_{\chi_1} < 2~\text{GeV}$. However, compared to the BaBar mono-$\gamma$ bound, the improvement is less pronounced when $\Delta_{\chi} = 0.1M_{\chi_1}$. This is because a smaller mass splitting leads to a longer $\chi_2$ lifetime, reducing the detection efficiency at Belle II. 
For $\Delta_{\chi} = 0.2M_{\chi_1}$, near $M_{\chi_1} = 1.8$ GeV, $\epsilon$ can reach as low as $2 \times 10^{-4}$, resulting in an exclusion region that significantly exceeds the BaBar mono-$\gamma$ constraints. As expected, mono-photon limits dominate in the regime of small mass splitting $\Delta_{\chi}$ as the primary discovery channel, while dark Higgs-strahlung process identifies the underlying model. However, displaced vertex searches surpass mono-photon sensitivity at larger $\Delta_{\chi}$. To verify this statement, we adopt the model parameters $\alpha_D = 0.1$, $m_{h_2} = m_{A'} = 3M_{\chi_1}$, and $\Delta_{\chi} = 0.4M_{\chi_1}$ to display results in Fig.~\ref{fig:bounds_04} of Appendix~\ref{appendix:C}. Therefore, these two future bounds are complementary and should be considered simultaneously.

For the rare $B$ meson decay process, an optimistic integrated luminosity value of 50 $ab^{-1}$ with the selection criteria in Table~\ref{tab:cut_flow_1} for the fermionic inelastic DM models at Belle II is applied, and we predict future bounds, as shown in Fig.~\ref{fig:bounds_2}. We fix the parameters $\alpha_D = 10^{-3}$, $m_{h_2} = 2.5M_{\chi_1}$ ,$m_{A^{\prime}} = 2.01M_{\chi_1} + \Delta_{\chi}$, and $\Delta_{\chi} = 0.1M_{\chi_1}$ (blue region) or $\Delta_{\chi} = 0.2M_{\chi_1}$ (red region), applying 90\% C.L. contours corresponding to an upper limit of 2.3 events under a background-free assumption.

Compared to the dark Higgs-strahlung process, for the same $M_{\chi_1}$, $\Delta_\chi$, and $\epsilon$, the rare $B$ meson exotic decay process results in a larger displacement between the displaced vertex and the IP because of the much smaller $\alpha_D$. This causes the lower boundary of the constrained parameter space to shift upward. Another key distinction is that the cross-section of the rare $B$ meson decay process is independent of $\epsilon$, making the lower boundary of the constrained parameter space steeper. This also results in an upper boundary of the parameter space with $\epsilon > 0.1$.
In the case of $\Delta_\chi = 0.1M_{\chi_1}$, the flight distance of $\chi_2$ at low masses is too long, requiring an increase in $\epsilon$ to enhance its decay width. As a result, the left boundary of the excluded parameter space is restricted to $M_{\chi_2} > 0.4$ GeV.
Similar to the dark Higgs-strahlung process, the parameter space covers the smallest $\epsilon$ at $M_{\chi_1} \approx 1.8$ GeV. For $\Delta_\chi = 0.1M_{\chi_1}$, unfortunately, the Belle II exclusion region is entirely contained within the BaBar and Belle II mono-$\gamma$ bounds. However, for $\Delta_\chi = 0.2M_{\chi_1}$, the exclusion region significantly surpasses the BaBar mono-$\gamma$ and LEP bounds. Most importantly, it can further explore the close-to-resonance~\cite{Berlin:2017ftj} annihilation scenario in the range $1~\text{GeV} < M_{\chi_1} < 2~\text{GeV}$. Moreover, the projected mono-$\gamma$ bounds at Belle II with 50 $ab^{-1}$ can further cover almost all of the predicted relic abundance in Fig.~\ref{fig:bounds_2}. Additionally, due to the absence of published analyses for NuCal, CHARM, and LSND constraints near the resonant DM co-annihilation regions, we conservatively exclude them here.

\section{Dark Higgs boson invisible decays at Belle II and $\chi_2$ very long-lived decays in GAZELLE detector}
\label{sec5}

In this section, we first discuss how the invisible decay of $h_2$ in the dark Higgs-strahlung process affects the previously studied $e^+ e^-\rightarrow \chi_1\chi_2$ process and investigate whether the invisible decay of $h_2$ in the rare $B$ meson decay process can account for the recent $B\rightarrow K\nu\bar{\nu}$ excess observed at Belle II~\cite{Belle-II:2023esi}. Additionally, we explore the potential for detecting semi-visible dark Higgs boson signatures with the proposed far detector related to Belle II, GAZELLE~\cite{Dreyer:2021aqd}.

\subsection{Invisible decay of $h_2$ in dark Higgs-strahlung and rare $B$
meson decay processes}

In this subsection, we compare the kinematic distributions of the displaced vertices for the $e^+ e^- \rightarrow \chi_1 \chi_2\rightarrow \chi_1\chi_1 e^+ e^-$ process and the invisible decay of $h_2$ in the dark Higgs-strahlung process, $e^{+} e^{-}\rightarrow h_{2} A^{\prime} \rightarrow (\chi_1 \chi_1) (\chi_2 \chi_1) \rightarrow (\chi_1 \chi_1) (\chi_1 e^{+} e^{-} \chi_1)$, as shown in Fig.~\ref{fig:DM_kineDis_11} for BP-1 and BP-2 and Fig.~\ref{fig:DM_kineDis_12} for BP-3 and BP-4 of Appendix~\ref{appendix:C}. 
A common feature of these two processes is the presence of only one displaced vertex. 
It should be noted that we compare the kinematic distributions of the $e^+ e^- \rightarrow \chi_1 \chi_2\rightarrow \chi_1\chi_1 e^+ e^-$ process alone (dashed lines) and the scenario where both processes occur simultaneously (solid lines), without normalization.

The process $e^+ e^- \rightarrow \chi_1 \chi_2$ produces $\chi_2$ directly from the $e^+ e^-$ collision, whereas in the invisible decay of $h_2$ in the dark Higgs-strahlung process, $\chi_2$ is produced via $A'$, which itself originates from the $e^+ e^-$ collision. As a result, $\chi_2$ from the former process is more likely to be boosted, leading to higher-energy electrons from its decay. This trend is evident in Fig.~\ref{fig:DM_kineDis_11}, where the electrons from dark Higgs-strahlung process only distribute in the range $E_{e^-} < 0.4$ GeV for BP-1 and BP-2 and $E_{e^-} < 0.8$ GeV for BP-3 and BP-4 when both processes produce a pair of electrons simultaneously. 
Additionally, the energy contribution from the electron pair in both processes is primarily within $E_{e^+e^-} < 0.6$ GeV for BP-1 and BP-2 and $E_{e^+e^-} < 1.2$ GeV for BP-3 and BP-4, whereas in the $e^+ e^- \rightarrow \chi_1 \chi_2$ process, the electron pair energy can reach up to 1.1 GeV for BP-1 and BP-2 and 2 GeV for BP-3 and BP-4. Since $\chi_2$ from the dark Higgs-strahlung process is less likely to be boosted, the angular separation between a pair of electrons from its decay is concentrated in the region $\Delta R_{e^+e^-} > 0.2$ for BP-1 and BP-3 and $\Delta R_{e^+e^-} > 0.4$ for BP-2 and BP-4, as larger $M_{\chi_2}$ results in a larger $\Delta R_{e^+e^-}$. 
For the $r_z$ distribution, in the dark Higgs-strahlung process, since $\chi_2$ is produced via $A'$, it is less boosted. This results in a smaller $r_z$ distribution for the displaced vertex, particularly when $M_{\chi_1}$ and $\Delta_\chi$ are larger.

For these two processes, if the invisible decay of $h_2$ in the dark Higgs-strahlung process is not considered, 
the cross-sections for BP-1 to BP-4 would decrease by $40.85\%$, $37.99\%$, $49.44\%$, and $45.40\%$, respectively, at the generator level. 
This reduction primarily impacts the experimentally observable distributions in regions with lower-energy displaced vertices located closer to the IP, as well as in regions where the displaced vertices produce lower-energy electrons and larger opening angles between electron pairs.

Additionally, according to Ref.~\cite{Belle-II:2023esi}, the current observed branching ratio for $B\to K\nu\bar{\nu}$ is $(2.3\pm 0.7)\times 10^{-5}$ which is a mild $2.7\sigma$ excess compared to the SM prediction. To account for contributions from dark Higgs boson invisible decays, we consider the $2\sigma$ allowed region. Since the SM contribution to this rare process is $BR(B\to K\nu\bar{\nu}) = (0.497\pm 0.037)\times 10^{-5}$~\cite{Ciuchini:2016weo}, the additional contribution from dark Higgs decays must be determined after subtracting the SM contribution. In the inelastic DM models considered here, there are two ways to mimic the $B\to K\nu\bar{\nu}$ signature. The first possibility is through the decay chain $B \rightarrow K h_2, h_2 \rightarrow \chi_1 \chi_1 $ and the second is via $B \rightarrow K h_2, h_2 \rightarrow \chi_2 \chi_2$ with the requirement that $\chi_2$ is sufficiently long-lived to escape the detector and thus remain undetected at Belle II.

\begin{figure}[t!]\centering
\includegraphics[width=0.6\textwidth]{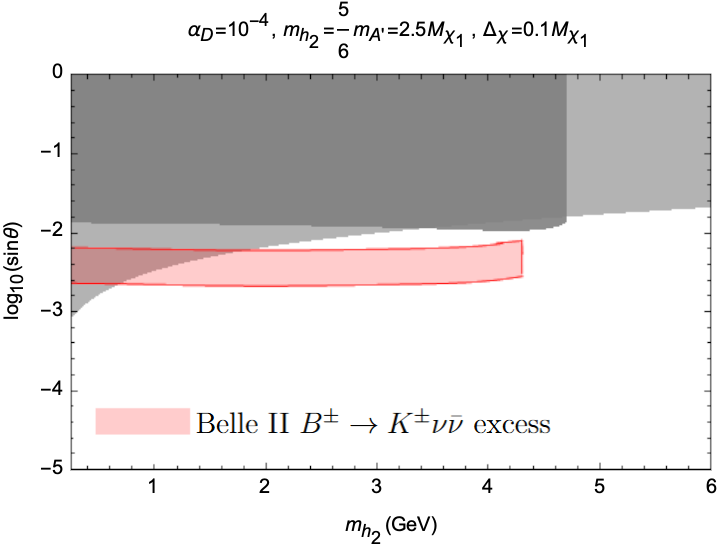}
\caption{
In the $ (m_{h_2} \text{ (GeV)}, \log_{10} \sin\theta) $ plane, we display a benchmark scenario with $\alpha_D=10^{-4}$ and $\Delta_{\chi} =0.1M_{\chi_1}$, where the mass spectrum is fixed as $ m_{h_2} = \frac{5}{6} m_{A'} = 2.5 M_{\chi_1} $. The gray region denotes the parameter space excluded by existing constraints on $h_2$ (as shown in Fig.~\ref{fig:case1}), while the red region represents the parameter space where the invisible decays of the dark Higgs boson account for the observed excess in $B^{\pm}\to K^{\pm}\nu\bar{\nu}$ at Belle II.
}
\label{fig:Br_Kh2}
\end{figure}

Considering eight parameter sets from Fig.~\ref{fig:case1} and Fig.~\ref{fig:case2} of Appendix~\ref{appendix:C}, we compute the contribution of invisible dark Higgs boson decays to the rare $B$ meson decay branching ratio beyond the SM. As shown in Fig.~\ref{fig:h2 decay ratio}, the sum of the branching ratios for $h_2$ decaying into $\chi_1\chi_1$ and $\chi_2\chi_2$ is close to 1. In light of current experimental constraints, we impose the condition $\epsilon < 5 \times 10^{-4}$. Referring to the method in Ref.~\cite{Lu:2023cet}, we compute $c\tau_{\chi_2}$ and find that the probability of $\chi_2$ escaping the Belle II detector is nearly 1. 
Under the eight parameter sets considered in Fig.~\ref{fig:case1} and Fig.~\ref{fig:case2} of Appendix~\ref{appendix:C}, the constrained regions are nearly identical. Thus, we present only the parameter settings corresponding to the top-left panel of Fig.~\ref{fig:case1}. The final constrained region is displayed in Fig.~\ref{fig:Br_Kh2}. 
The gray region represents the parameter space excluded by current constraints on $h_2$ (as shown in Fig.~\ref{fig:case1}), while the red region corresponds to the parameter space where invisible decays of the dark Higgs boson account for the observed $B^{\pm}\to K^{\pm}\nu\bar{\nu}$ excess at Belle II. 
As illustrated in Fig.~\ref{fig:Br_Kh2}, for $ m_{h_2} > 0.6 $ GeV, invisible decays of the dark Higgs boson remain viable to explain this anomaly. Moreover, because sizable values of $\sin\theta$ (within $2.51\times 10^{-3}$ to $6.31\times 10^{-3}$) are required, the values of $\alpha_D$ must be kept small. Consequently, if DM annihilation in the early universe occurs predominantly via co-annihlation, a close-to-resonance scenario is necessary to match the observed relic abundance.

\subsection{Search for semi-visible dark Higgs boson in the L-GAZELLE detector}

In this subsection, we employ the L-GAZELLE detector to search for dark Higgs-strahlung and rare $B$ meson decay processes, addressing the limitations of Belle II in detecting long-lived $\chi_2$ particles. The GAZELLE detector enhances the sensitivity of Belle II and the International Linear Collider (ILC) to LLPs by enabling distant tracking and displaced vertex reconstruction~\cite{Dreyer:2021aqd}. At Belle II, GAZELLE is designed with three configurations to optimize LLP detection across different angular and radial ranges~\cite{Dreyer:2021aqd}: 
\begin{itemize}
    \item \textbf{Baby-GAZELLE (BG)}: A cubic detector with dimensions $4~\text{m} \times 4~\text{m} \times 4~\text{m}$, located on the floor of Tsukuba Hall at approximately $x=10~\text{m}, y=-3.7~\text{m}, z=10~\text{m}$. It covers a solid angle of $\Omega = 0.12$ sr, corresponding to approximately 0.95\% of the total solid angle.
    
    \item \textbf{L-GAZELLE (LG)}: Comprising two components:
        \begin{itemize}
            \item \textbf{LG-B1}: A detector with dimensions $6~\text{m} \times 16~\text{m} \times 24~\text{m}$, positioned at approximately $x=35~\text{m}, y=2.3~\text{m}, z=0~\text{m}$, covering a solid angle of $\Omega = 0.34$ sr (2.7\% coverage).
            \item \textbf{LG-B2}: A detector with dimensions $26~\text{m} \times 16~\text{m} \times 3~\text{m}$, positioned at approximately $x=19~\text{m}, y=2.3~\text{m}, z=10.5~\text{m}$, covering a solid angle of $\Omega = 0.76$ sr (6.0\% coverage).
        \end{itemize}
    
    \item \textbf{GODZILLA (GZ)}: A detector with dimensions $25~\text{m} \times 10~\text{m} \times 50~\text{m}$, located outside Tsukuba Hall at approximately $x=-27~\text{m}, y=18~\text{m}, z=20~\text{m}$. It covers a solid angle of $\Omega = 0.74$ sr, corresponding to approximately 5.9\% of the total solid angle.
\end{itemize}

Due to its wide angular coverage and moderate detection distance, L-GAZELLE is well-suited to study the long-lived $\chi_2$ generated from dark Higgs-strahlung and rare $B$ meson exotic decay processes, making it the optimal choice for this analysis. The cleaner experimental environment and lower background noise of the Belle II experiment, combined with GAZELLE's capabilities, such as precise particle direction and timing information, enable effective background event rejection. Consequently, a background-free scenario is considered in this part of the analysis.

We demonstrate our analysis in the parameter space of $(M_{\chi_1}, \epsilon)$, considering the same two signal processes and parameter settings as in the previous analysis as shown in Fig.~\ref{fig:bounds} and Fig.~\ref{fig:bounds_2}, respectively. Here, only the basic event selections are employed. At the generator level, we required $p^{\ell}_T > 0.01$ GeV, $|\eta_{\ell}|<2.5$ and $\Delta R_{\ell\ell}>0.01$ for charged leptons. To simulate the L-GAZELLE detector, we use the software {\tt Displaced Decay Counter} (DDC)~\cite{Ferber:2023iso} to evaluate the detector’s sensitivity, assuming that an LLP is detected whenever and wherever it decays within the detector volume. 

The DDC program estimates the probability of a given LLP decaying within the corresponding detector. 
Specifically, along its trajectory, the probability that the particle decays after traveling a distance between $\ell_1$ and $\ell_2$ in the lab frame is given by
\begin{align}
\mathcal{P}(\ell_{1},\ell_{2}) = \exp\left[-\frac{\ell_{1}}{\gamma\beta\tau}\right] - \exp\left[-\frac{\ell_{2}}{\gamma\beta\tau}\right],
\label{eq:L-GAZELLE}
\end{align}
where $\beta$ is the particle's velocity, $\gamma = (1 - \beta^2)^{-1/2}$ is the Lorentz factor, and $\tau$ is the particle's lifetime. 
We construct the L-GAZELLE detector using the DDC program. Here, \( \ell_1 \) is the distance between the particle and the IP when it enters the detector, and \( \ell_2 \) is the distance when it exits. Given the detector geometry, \( \ell_1 \) and \( \ell_2 \) primarily depend on the particle's $\eta$.

We simulate the L-GAZELLE detector using its geometric information by the DDC program, which calculates the probability of an LLP decaying inside L-GAZELLE by determining $\ell_1$ and $\ell_2$ for each LLP based on its flight direction, lifetime, Lorentz factor, and the detector's boundaries along the LLP’s flight path.
Combining the particle production cross-section (\(\sigma\)) and the visible branching ratio (\(BR_{\text{visible}}\)), the number of signal events (\(N_{\text{events}}\)) that the detector is expected to detect under a specific integrated luminosity  (\(\mathcal{L}\)) is calculated:
\[ N_{\text{events}} = \mathcal{L} \times \sigma \times BR_{\text{visible}} \times \mathcal{P}(\ell_{1},\ell_{2}). \] 
For instance, if a LLP travels horizontally, it may decay within the LG-B1 component, with corresponding values of $l_1 = 31.9$ m and $l_2 = 37.9$ m. For BP-A to BP-D, the average values of $\gamma \beta$ are about $1.51$, $0.63$, $1.34$, and $0.42$, respectively. The probability of an LLP traveling horizontally and decaying within L-GAZELLE can be calculated using Eq.~(\ref{eq:L-GAZELLE}), yielding probabilities of $0.40\%$, $5.65\%$, $3.05\%$, and $9.88 \times 10^{-10}$, respectively. 
Note that Eq.~(\ref{eq:L-GAZELLE}) calculates the probability for one LLP to decay within the far detector. In our analysis, both displaced vertices from two the $\chi_2$ particles must be detected. Since both vertices originate from the decay of $\chi_2$ following the decay of $h_2$, they exhibit similar spatial distributions. 
For BP-A, after applying the basic selection criteria, we use the DDC program to calculate the detection efficiency. The efficiency is $1.59\%$ when requiring at least one displaced vertex within the detector, and $1.33\%$ when requiring two displaced vertices, a reduction of $15.2\%$. For BP-B to BP-C, the reduction in efficiency when requiring two displaced vertices instead of one does not exceed $24\%$. Therefore, to ensure background-free conditions and enable signal identification in the L-GAZELLE analysis, we require detection of both displaced vertices from $\chi_2$ decays within the detector volume.

Finally, we apply an optimistic integrated luminosity value of 50 $ab^{-1}$ with only basic event selections at the generator
level for fermionic inelastic DM models at L-GAZELLE detector, and predict future bounds (dashed lines) for the dark Higgs-strahlung and rare $B$ meson exotic decay processes in Fig.~\ref{fig:bounds} and Fig.~\ref{fig:bounds_2}, respectively. We use the same parameter settings for these two processes as before, with $\Delta_{\chi} = 0.1M_{\chi_1}$ (blue region) and $\Delta_{\chi} = 0.2M_{\chi_1}$ (red region), and apply 90\% C.L. contours corresponding to an upper limit of 2.3 events under the assumption of a background-free scenario. We observe that L-GAZELLE complements Belle II’s limited efficiency for detecting LLPs with extended flight distances. A smaller $M_{\chi_2}$, smaller $\epsilon$, or smaller $\Delta_\chi$ leads to a longer $\chi_2$ lifetime, shifting the displaced vertex beyond the reach of Belle II's detector. As a far detector, L-GAZELLE covers regions with smaller $\epsilon$ and $M_{\chi_1}$.

For the dark Higgs-strahlung process, the cross-section is proportional to $\epsilon^2$. The increasing $M_{\chi_2}$ reduces  the flight distance and leads to a decrease in $\epsilon$, extending the exclusion region to lower values. When $M_{\chi_2}$ exceeds approximately 1 GeV, the flight distance shortens further, causing the upper boundary of the exclusion region to drop more rapidly. 
Meanwhile, L-GAZELLE's detection efficiency for displaced vertices decreases as the flight distance shortens, requiring a larger $\epsilon$ to maintain a sufficient $N_{\text{events}}$. This raises the lower boundary and narrows the exclusion region. Ultimately, the right boundary extends to cover $M_{\chi_1} = 2$ GeV. 
For $\Delta_\chi = 0.2 M_{\chi_1}$, the shorter $\chi_2$ lifetime accelerates this narrowing effect, and the right boundary extends to cover $M_{\chi_1} = 1.7$ GeV.

For the rare $B$ meson exotic decay process, the cross-section is independent from $\epsilon$. This implies that for $M_{\chi_1} < 1$ GeV, the upper and lower boundaries of the constrained parameter space descend at a similar rate. For smaller $\Delta_\chi$, $\chi_2$ travels farther, so the upper and lower boundaries for $\Delta_\chi = 0.1 M_{\chi_1}$ are both higher than those for $\Delta_\chi = 0.2 M_{\chi_1}$. 
When $M_{\chi_1} > 1$ GeV, the reduction in cross-section and the lower detection efficiency of L-GAZELLE cause the constrained region's boundaries to gradually converge. Near $M_{\chi_1} \approx 1$ GeV, the lower boundary exhibits a notable inflection point. Ultimately, for $\Delta_\chi = 0.1 M_{\chi_1}$, the right boundary extends to $M_{\chi_1} = 1.8$ GeV, and for $\Delta_\chi = 0.2 M_{\chi_1}$, it reaches $M_{\chi_1} = 2$ GeV.

For the dark Higgs-strahlung process in Fig.~\ref{fig:bounds}, L-GAZELLE's exclusion limits exceed those of NuCal, CHARM and LSND bounds, LEP bounds, and BaBar mono-$\gamma$ bounds for both $\Delta_\chi = 0.1 M_{\chi_1}$ and $\Delta_\chi = 0.2 M_{\chi_1}$. For $\Delta_\chi = 0.1 M_{\chi_1}$, it surpasses BaBar mono-$\gamma$ bounds in the range $0.15$ GeV $< M_{\chi_1} < 1.1$ GeV and exceeds NuCal, CHARM, and LSND bounds in the range $0.8$ GeV $< M_{\chi_1} < 2$ GeV, with $\epsilon$ reaching $4 \times 10^{-4}$ near $M_{\chi_1} = 1$ GeV. For $\Delta_\chi = 0.2 M_{\chi_1}$, it exceeds BaBar mono-$\gamma$ bounds across all $M_{\chi_1}$ values and surpasses NuCal, CHARM, and LSND bounds in the range $0.9$ GeV $< M_{\chi_1} < 1.8$ GeV, reaching the parameter space minimum at $\epsilon = 2.5 \times 10^{-4}$ when $M_{\chi_1} = 1.1$ GeV.

Comparing Fig.~\ref{fig:bounds} and Fig.~\ref{fig:bounds_2}, the $\chi_2$ from rare $B$ meson decay travels farther, allowing L-GAZELLE to surpass more of the current LEP constraints. For $\Delta_\chi = 0.1 M_{\chi_1}$, in the range $0.2$ GeV $< M_{\chi_1} < 1.6$ GeV, it exceeds BaBar mono-$\gamma$ bounds. Near $M_{\chi_1} = 1$ GeV, $\epsilon$ can be as small as $1.2 \times 10^{-4}$, enabling exploration of the close-to-resonance~\cite{Berlin:2017ftj} annihilation scenario. For $\Delta_\chi = 0.2 M_{\chi_1}$, across the considered $M_{\chi_1}$ range, it consistently exceeds BaBar mono-$\gamma$ bounds. In the range $0.4$ GeV $< M_{\chi_1} < 1.8$ GeV, it covers regions below close-to-resonance annihilation scenario. Finally, at $M_{\chi_1} = 1.1$ GeV, it reaches the lowest point in the parameter space with $\epsilon = 5 \times 10^{-5}$. Again, the projected mono-photon limits from Belle II are quite strong and provide competitively complementary constraints to those of L-GAZELLE.

\section{Conclusion}
\label{sec6}

In this study, we explored the elusive dark Higgs boson within the framework of spin-1/2 inelastic dark matter (DM) models in the Belle II experiment. The dark Higgs boson plays a pivotal role in these models by generating the mass of the dark photon as well as inducing the mass splitting between the excited and ground states of DM. In particular dark Higgs boson plays a crucial role to restore unitarity in 
the Compton scattering ($\chi_1 A'_L \rightarrow \chi_1 A'_L$) or 
DM pair annihilation ($\chi_1 \chi_1 \rightarrow A'_L A'_L$). Therefore, it is important to search for dark Higgs boson in addition to DM and massive dark photon.
However, detecting such dark Higgs boson is challenging, particularly when its mass exceeds twice the mass of the DM excited state, resulting in semi-visible or invisible decay signatures. Our analysis focused on fermionic inelastic DM models with $U(1)_D$ dark gauge symmetry. We investigated two specific dark Higgs boson production processes at Belle II: dark Higgs-strahlung and rare $B$-meson exotic decays as shown in Fig.~\ref{fig:Feyn_diagram}. The inclusive two-dilepton displaced vertices and the associated missing energy produced by the dark Higgs bosn serve as significant indicators of its presence.

Under experimental constraints on the light dark Higgs boson in Fig.~\ref{fig:case1}, we specifically focus on two distinct parameter spaces. The first type features larger $\alpha_D$ and smaller $\sin\theta$, making it suitable for the dark Higgs-strahlung process ($e^+e^- \rightarrow A' h_2$). This process is highly sensitive to the combination of $\alpha_D$ and the kinetic mixing parameter $\epsilon$. A larger $\alpha_D$ enhances the production of the dark photon $A'$, increasing the yield of $h_2$. 
The second type of parameter space involves smaller $\alpha_D$ and moderate $\sin\theta$, which is more suitable for the rare $B$-meson exotic decay process ($B \rightarrow K h_2$). This process is primarily sensitive to $\sin\theta$ and less dependent on $\alpha_D$. A smaller $\alpha_D$ avoids conflicts with Higgs signal measurements and other experimental constraints, while a moderate $\sin\theta$ still allows for observable signal rates.
The complementary nature of these two processes enables broader coverage of the parameter space and a more effective utilization of Belle II's detection capabilities.

The event selection criteria in Table~\ref{tab:cut_flow_1} are applied to identify at least two dilepton displaced vertices at Belle II based on the kinematic distributions in Fig.~\ref{fig:DM_kineDis_1} and Fig.~\ref{fig:DM_kineDis_3}. 
Our results indicate that detection efficiency is strongly correlated with the spatial distribution of $\chi_2$ decay vertices and the mass splitting $\Delta_\chi$. 
Additionally, given the limitations of the Belle II detector in probing regions of parameter space where these $\chi_2$ have long lifetimes, we explored the potential of the proposed L-GAZELLE detector for detecting two dilepton displaced vertices signature. With its expanded angular coverage and moderate detection distance, L-GAZELLE is expected to complement Belle II by enhancing sensitivity to long-lived particles and extending coverage to a broader region of parameter space.

The key results from our analysis are highlighted in Fig.~\ref{fig:bounds} and Fig.~\ref{fig:bounds_2}. Firstly, Fig.~\ref{fig:bounds} presents the projected constraints for the dark Higgs-strahlung process, demonstrating that Belle II and the proposed L-GAZELLE detector can significantly extend beyond current experimental limits, particularly in the mass range $0.5 \, \text{GeV} < M_{\chi_1} < 2 \, \text{GeV}$. For $\Delta \chi = 0.2 M_{\chi_1}$, the lower boundary of the parameter space can reach $\epsilon = 2 \times 10^{-4}$. 
Secondly, Fig.~\ref{fig:bounds_2} displays the future constraints for the rare $B$-meson exotic decay process, indicating that L-GAZELLE can surpass existing experimental bounds and probe the close-to-resonance annihilation scenario in the range $0.2 \, \text{GeV} < M_{\chi_1} < 2 \, \text{GeV}$. For $\Delta_{\chi} = 0.2 M_{\chi_1}$, the lower boundary of the parameter space can reach $\epsilon = 5 \times 10^{-5}$.

\appendix

\section{The $B \to K^{(\ast)}h_2$ decay branching ratio}
\label{appendix:A}

Here we summarize the formula for the $B \to K^{(\ast)}h_2$ decay branching ratio. We closely follow the formula for $B \to K^{(\ast)}h_2$ decay branching ratio from Ref.~\cite{Kachanovich:2021pvx}. First, the branching ratio for $B \to Kh_2$ can be written as 
\begin{equation}
\text{BR}(B \to Kh_2) = \frac{\tau_B}{32\pi m_B^2} |C_{h_2 sb}|^2 \left( \frac{m_B^2 - m_K^2}{m_b - m_s} \right)^2 f_0(m_{h_2}^2) \frac{\sqrt{\lambda(m_B^2, m_K^2, m_{h_2}^2)}}{2m_B},
\label{}
\end{equation}
where $\lambda(x, y, z) = x^2 + y^2 + z^2 - 2(xy + xz + yz)$, and the effective coupling $C_{h_2 sb}$ is
\begin{equation}
C_{h_2 sb} = -\frac{3 \sin \theta \lambda_t m_b m_t^2}{16\pi^2 v^3},
\end{equation}
with $\lambda_t = V_{tb} V_{ts}^\ast$, and $V_{ij}$ is the element of the CKM matrix. The scalar form factor $f_0(q^2)$ relates to the required scalar hadronic matrix element via
\begin{equation}
\langle K | \bar{s}b | B \rangle = \frac{m_B^2 - m_K^2}{m_b - m_s} f_0(q^2),
\end{equation}
where $q = p_B - p_K$. For this form factor, we use the QCD lattice result from Fig.~19 of Ref.~\cite{USQCD:2022mmc}.

Second, the branching ratio for $B \to K^\ast h_2$ can be written as
\begin{equation}
\text{BR}(B \to K^\ast h_2) = \frac{\tau_B}{32 \pi m_B^2} |C_{h_2 sb}|^2 \frac{A_0(m_{h_2})^2}{(m_b + m_s)^2} \frac{\lambda(m_B^2, m_{K^\ast}^2, m_{h_2}^2)^{3/2}}{2 m_B}.
\end{equation}
The form factor $A_0(q^2)$ relates to the required pseudoscalar hadronic matrix element via
\begin{equation}
\langle K^\ast(k, \epsilon) | \bar{s}\gamma_5 b | B(p_B) \rangle = \frac{2m_{K^\ast} \epsilon^\ast \cdot q}{m_b + m_s} A_0(q^2),
\end{equation}
where $\epsilon$ is the polarization vector of $K^\ast$, and $q = p_B - k$. For this form factor, we use the combination of QCD lattice and QCD sum rules from Fig.~2 of Ref.~\cite{Iwanaga:2016rne}.

\section{The $\Upsilon(1S) \to h_2\gamma$ exotic decay}
\label{appendix:B}

In this appendix, we focus on the anomalous decay of the $\Upsilon(1S)$ meson to $h_2\gamma$.
According to Ref.~\cite{Dong:2012hc}, the mass of the $\Upsilon(1S)$ is $M_{\Upsilon(1S)} = 9460.30$ MeV, and its width is $\Gamma_{\Upsilon(1S)} = 54.02$ keV. The branching ratio for $\Upsilon(1S)$ decaying to lepton pairs $l^+l^-$ is $7.46\%$, and the leptonic partial width can be expressed as
\begin{equation}
\Gamma_{\Upsilon(1S) \to l^+ l^-} = \frac{4\pi \alpha_{em}^2 F_{\Upsilon(1S)}^2}{27 M_{\Upsilon(1S)}} \left(1 + 2 \frac{m_l^2}{M_{\Upsilon(1S)}^2}\right) \sqrt{1 - 4 \frac{m_l^2}{M_{\Upsilon(1S)}^2}},
\label{1Sh2a}
\end{equation}
where $\alpha_{em}$ is the electromagnetic fine-structure constant, $m_l$ is the lepton mass, and $F_{\Upsilon(1S)}$ is the decay constant of the $\Upsilon(1S)$ meson, defined by the matrix element of the electromagnetic current
\begin{equation}
\langle 0 | \bar{b}\gamma_\mu b | \Upsilon(P, \epsilon) \rangle = F_{\Upsilon(1S)} M_{\Upsilon(1S)} \epsilon_\mu.
\end{equation}
In the non-relativistic limit, the decay constant is given by
\begin{equation}
F_{NR}^{\Upsilon(1S)} = \sqrt{\frac{12}{M_{\Upsilon(1S)}} |\Psi_{\Upsilon(1S)}(0)|},
\end{equation}
where $\Psi_{\Upsilon(1S)}(0)$ is the non-relativistic wave function at the origin. 
Via the Wilczek mechanism~\cite{Wilczek:1977zn,Dreiner:2008tw}, the partial width for $\Upsilon(1S) \to h_2\gamma$ can be expressed as the following ratio:
\begin{equation}
R_0 \equiv \frac{\Gamma_{\Upsilon(1S) \to h_2\gamma}}{\Gamma_{\Upsilon(1S) \to l^+ l^-}} = \frac{G_F m_b^2}{\sqrt{2}\pi\alpha_{em}} \sin^2\theta \left(1 - \frac{m_{h_2}^2}{M_{\Upsilon(1S)}^2}\right) C,
\end{equation}
where $G_F$ is the Fermi constant, $m_b$ is the bottom quark mass ($M_{\Upsilon(1S)} \approx 2m_b$), and $C \sim 1/2$ includes QCD corrections~\cite{Karkaryan:2024mgp,Nason:1986tr}.

\section{Supplementary figures}
\label{appendix:C}

\begin{figure}[t!]\centering
\includegraphics[width=0.48\textwidth]{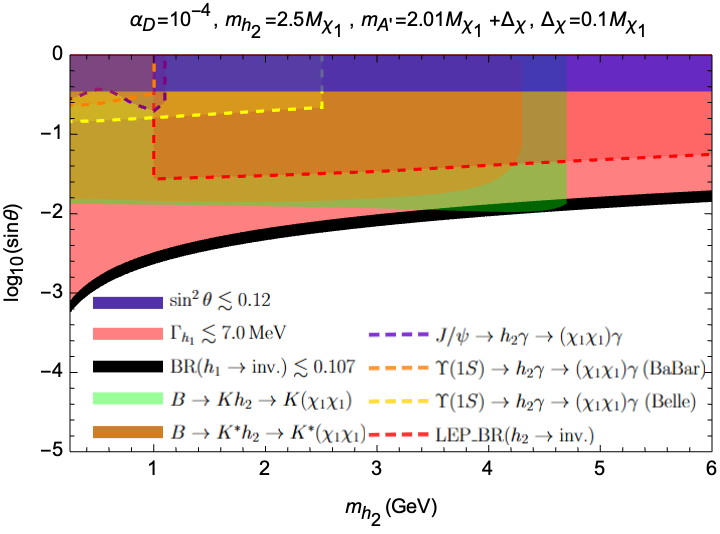}
\includegraphics[width=0.48\textwidth]{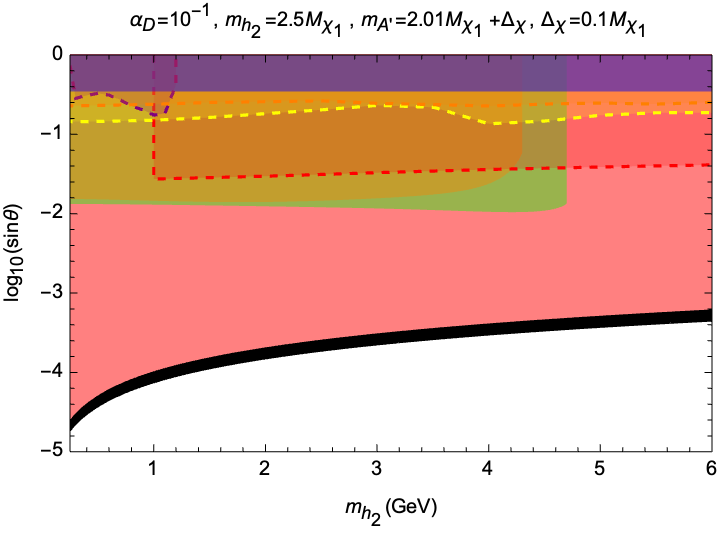}
\includegraphics[width=0.48\textwidth]{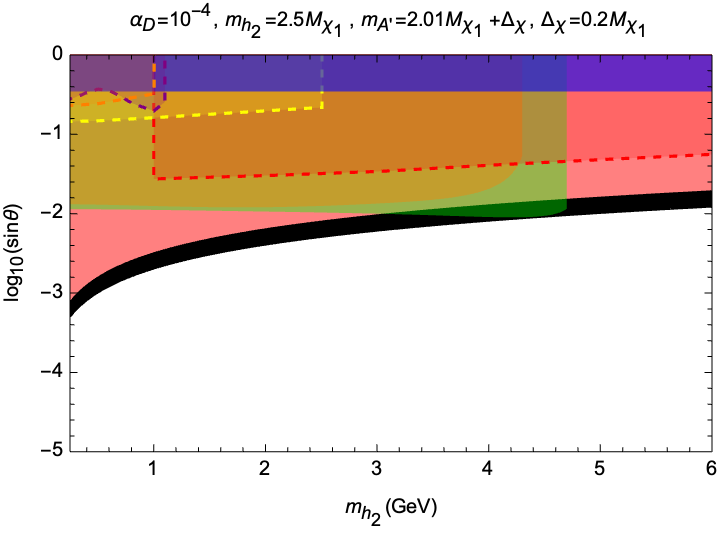}
\includegraphics[width=0.48\textwidth]{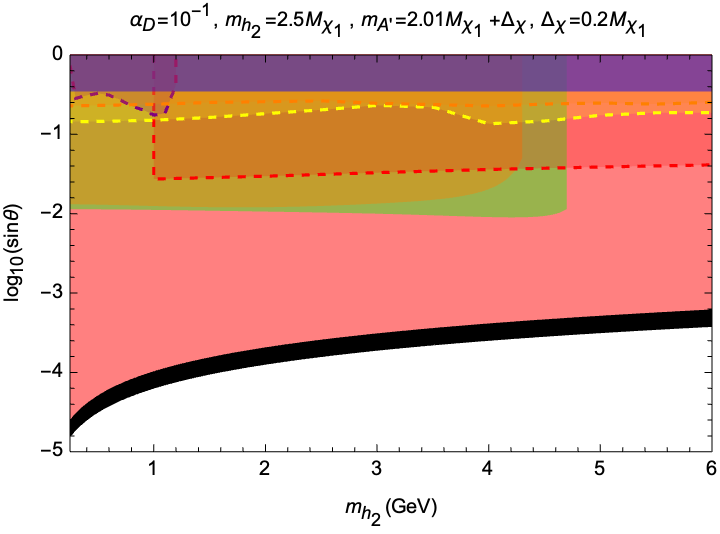}
\caption{ 
The relevant constraints for $ h_2 $ on the $ (m_{h_2} \text{ (GeV)}, \log_{10}\sin\theta) $ plane. We consider four benchmark points: $ (\alpha_D, \Delta_{\chi}/M_{\chi_1}) =$ 
$(10^{-4}, 0.1)$ (upper-left), $(10^{-1}, 0.1)$ (upper-right), $(10^{-4}, 0.2)$ (lower-left), $(10^{-1}, 0.2)$ (lower-right). Here, we have fixed the mass spectrum $ m_{h_2} = 2.5 M_{\chi_1} $ and $ m_{A'} = 2.01 M_{\chi_1} + \Delta_{\chi} $.  
}
\label{fig:case2}
\end{figure}

\begin{figure*}[ht!]
\centering{\includegraphics[width=0.48\textwidth]{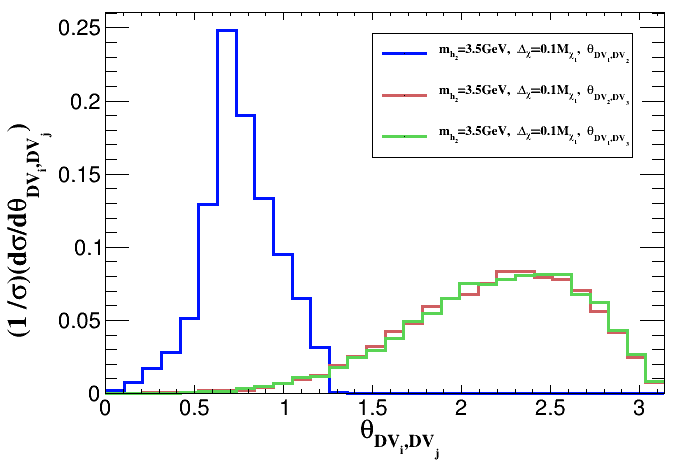}}
\centering{\includegraphics[width=0.48\textwidth]{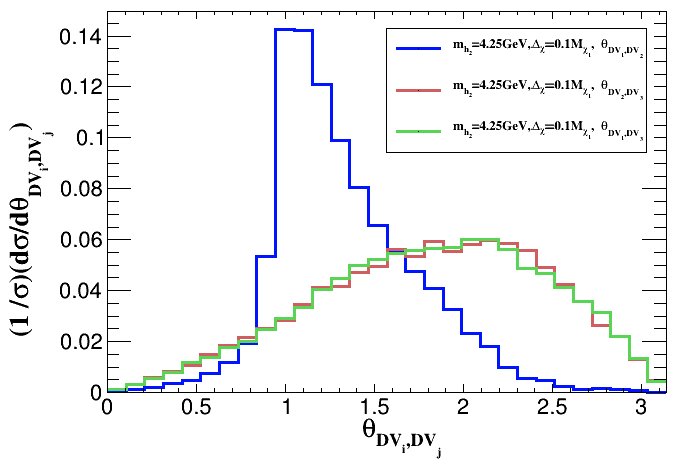}}
\centering{\includegraphics[width=0.48\textwidth]{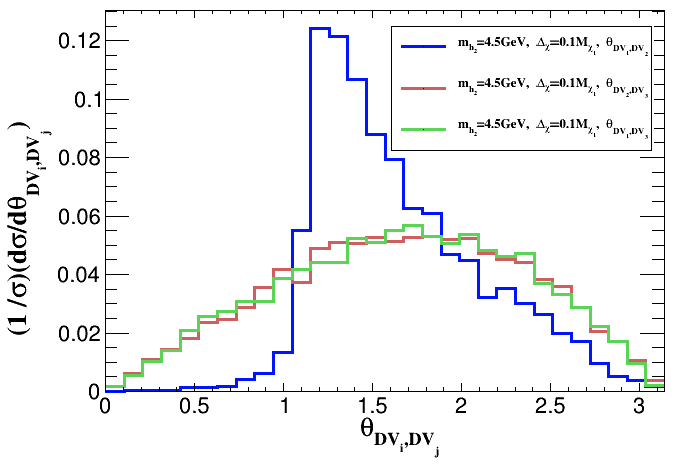}}
\centering{\includegraphics[width=0.48\textwidth]{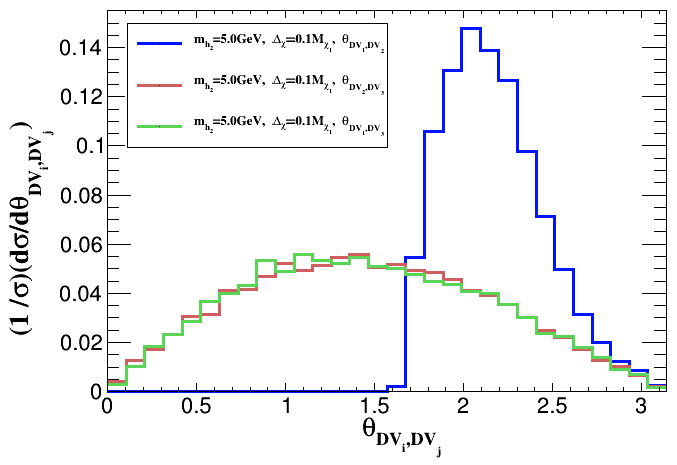}}
\caption{ Comparison between open angles of decay vertices: $\theta_{DV_{1},DV_{2}}$ and $\theta_{DV_{1},DV_{3}}$ at $M_{\chi_1} = 3.5$, $4.25$, $4.5$, and $5.0$ GeV, respectively.}
\label{fig:Kine_theta}
\end{figure*}

\begin{figure*}[ht!]
\centering{\includegraphics[width=0.9\textwidth]{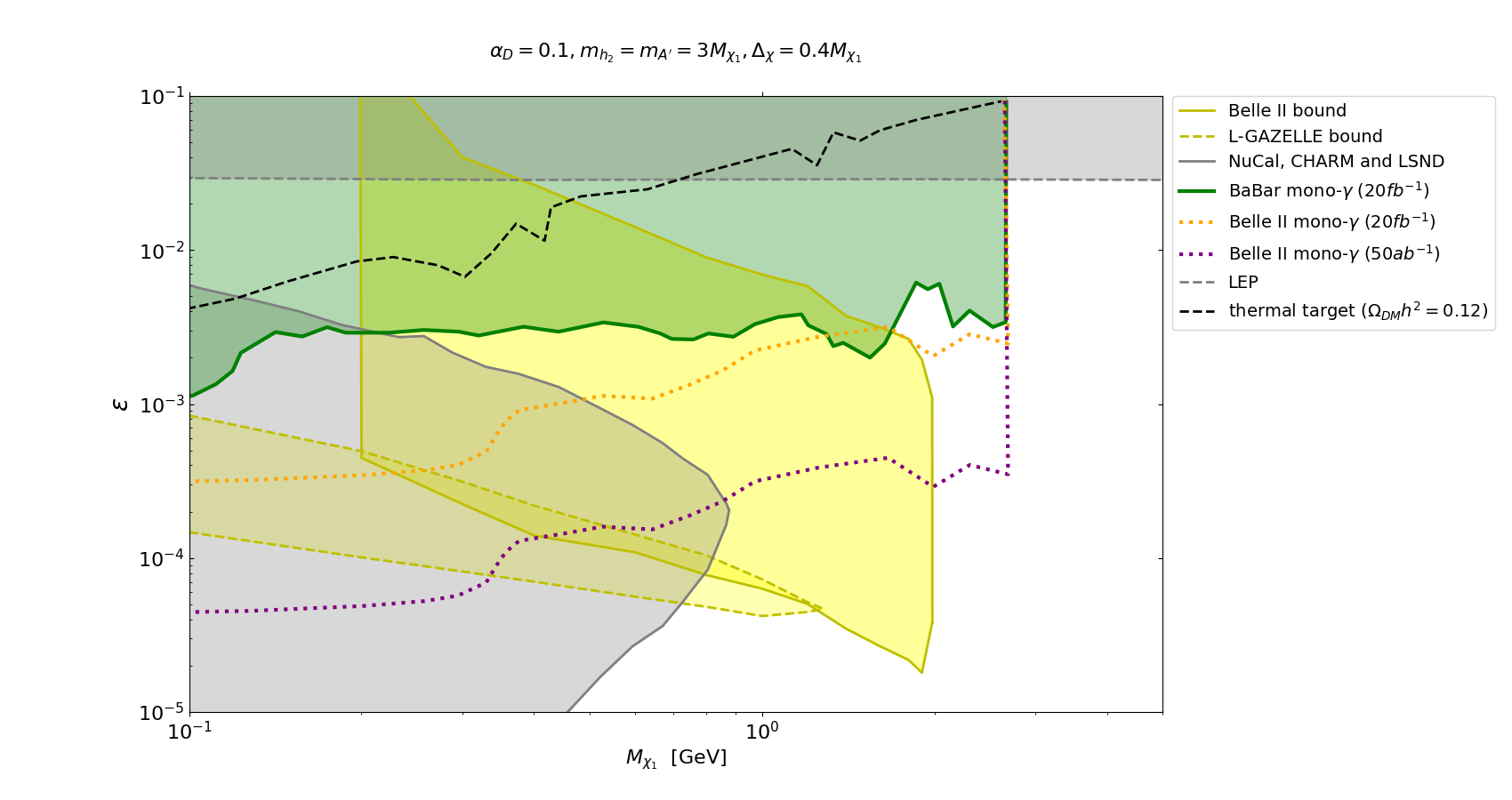}}
\caption{Same as Fig.~\ref{fig:bounds}, but for model parameters $\alpha_{D} = 0.1$, $m_{h_2} = m_{A^{\prime}} = 3M_{\chi_1}$ and $\Delta_{\chi} = 0.4M_{\chi_1}$ (yellow regions) are fixed and $90\%$ C.L. contours which correspond to an upper limit of 2.3 events with the assumption of background-free are applied. } 
\label{fig:bounds_04}
\end{figure*}

\begin{figure*}[ht!]
\centering{\includegraphics[width=0.42\textwidth]{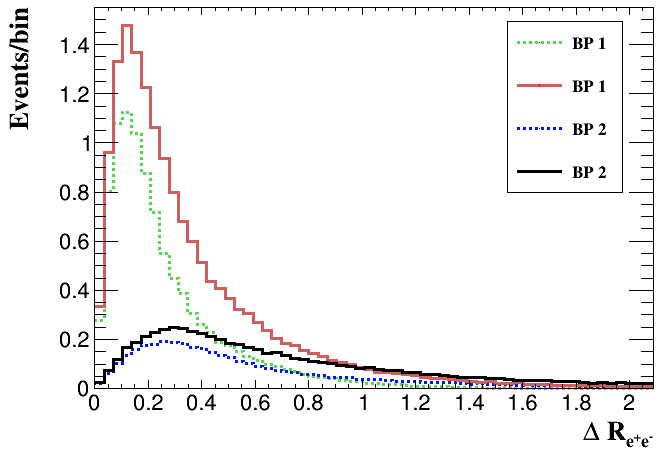}}
\centering{\includegraphics[width=0.42\textwidth]{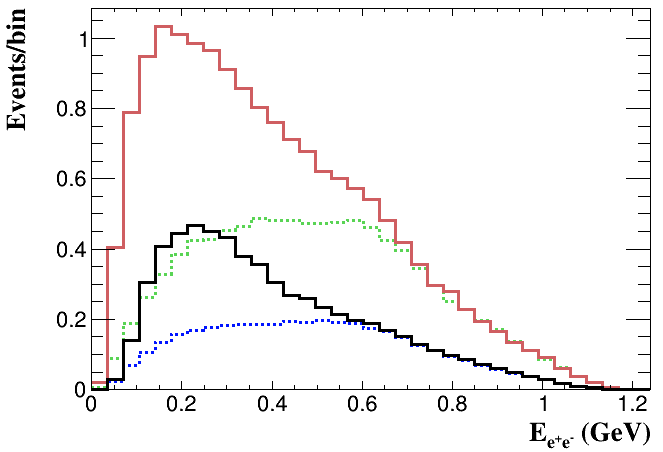}}
\centering{\includegraphics[width=0.42\textwidth]{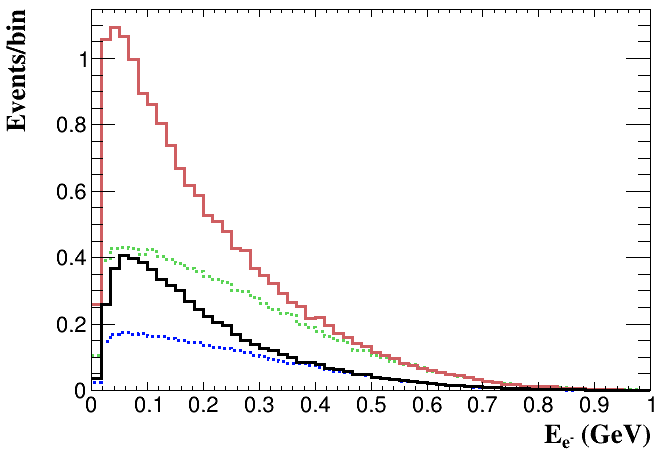}}
\centering{\includegraphics[width=0.42\textwidth]{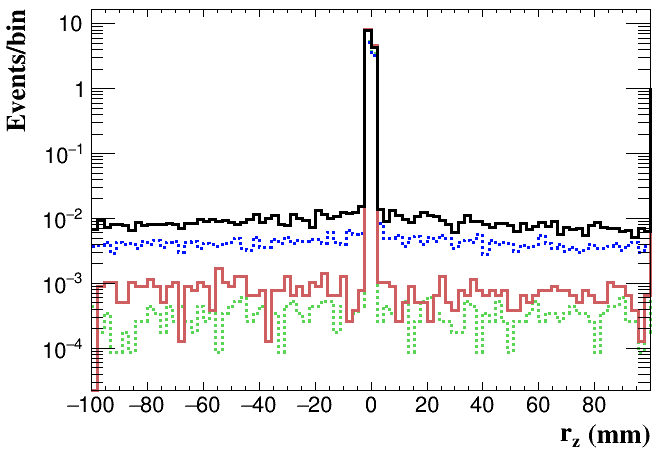}}
\caption{ Various kinematic distributions for process $e^{+} e^{-} \rightarrow \chi_2 \chi_1 \rightarrow (\chi_1 e^{+} e^{-} \chi_1)$ (dashed lines) and the invisible decay of $h_2$ in the dark Higgs-strahlung process (soiled lines) based on the BP-1 and BP-2 mentioned in the main text. In order, they are angular distance between electron pair, $\Delta R_{e^{+}e^{-}}$, electron pair energy, $E_{e^{+}e^{-}}$ (GeV), electron energy, $E_{e{-}}$ (GeV), longitudinal distribution of the decay vertices, $r_z$ (mm).}
\label{fig:DM_kineDis_11}
\end{figure*}

\begin{figure*}[ht!]
\centering{\includegraphics[width=0.42\textwidth]{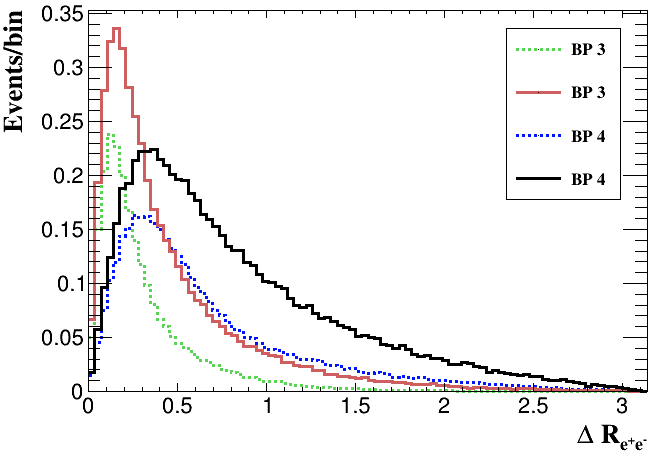}}
\centering{\includegraphics[width=0.42\textwidth]{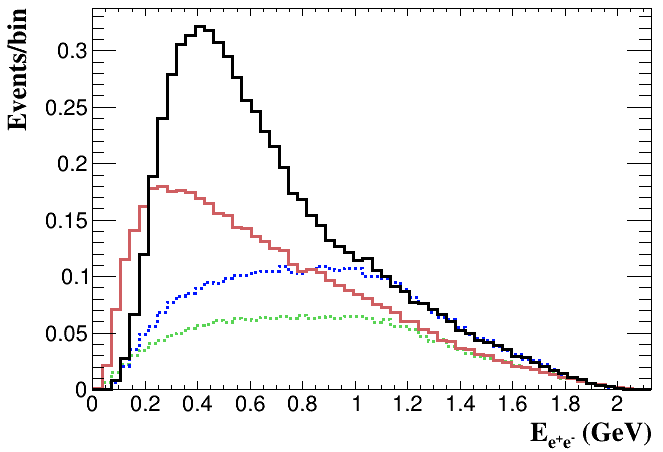}}
\centering{\includegraphics[width=0.42\textwidth]{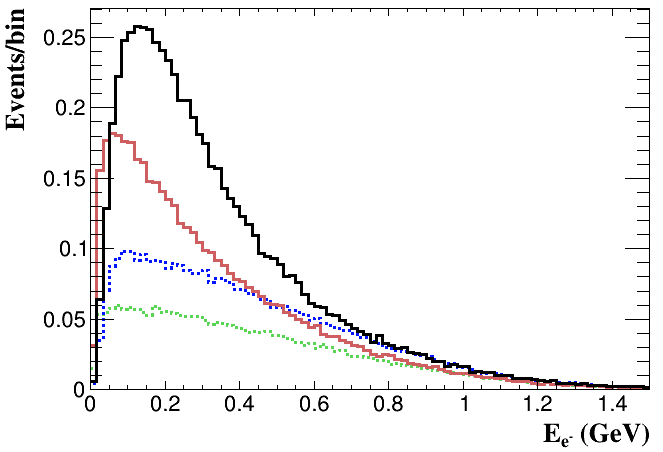}}
\centering{\includegraphics[width=0.42\textwidth]{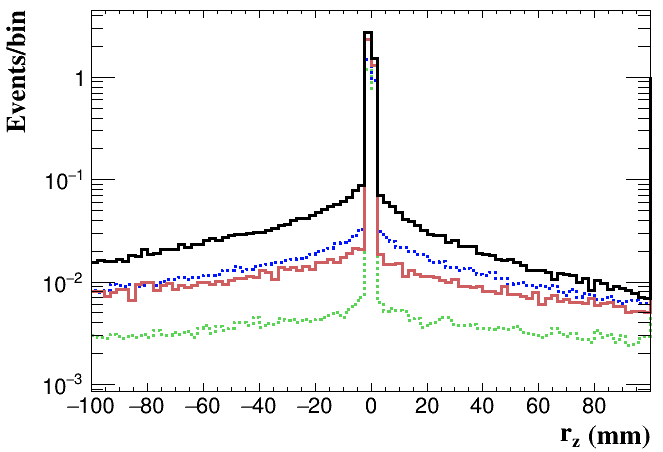}}
\caption{The same kinematic distributions as Fig.~\ref{fig:DM_kineDis_11}, but for BP-3 and BP-4.}
\label{fig:DM_kineDis_12}
\end{figure*}

In this Appendix, we present some additional figures which have been mentioned in the main text: 
\begin{itemize}

\item Similar to Fig.~\ref{fig:case1}, we further consider four identical benchmark points: $(\alpha_D, \Delta_{\chi}/M_{\chi_1}) = (10^{-4}, 0.1)$ (upper-left), $(10^{-1}, 0.1)$ (upper-right), $(10^{-4}, 0.2)$ (lower-left), and $(10^{-1}, 0.2)$ (lower-right), but for $m_{h_2} = 2.5M_{\chi_1}$ and $m_{A'} = 2.01M_{\chi_1} + \Delta_{\chi}$ in Fig.~\ref{fig:case2}. 

\item Fig.~\ref{fig:Kine_theta} displays the distributions of $\theta_{DV_{1},DV_{2}}$ and $\theta_{DV_{1},DV_{3}}$ in the dark Higgs-strahlung process. When $m_{h_2} > 4$ GeV, due to the overlap of $\theta_{DV_{1},DV_{2}}$ and $\theta_{DV_{1},DV_{3}}$ exceeding 30\%, the displaced vertex of $A'$ must be included in the event selections.

\item {Fig.~\ref{fig:bounds_04} resembles Fig.~\ref{fig:bounds} but adopts fixed model parameters $\alpha_{D} = 0.1$, $m_{h_2} = m_{A^{\prime}} = 3M_{\chi_1}$, and $\Delta_{\chi} = 0.4M_{\chi_1}$ (yellow regions). The $90\%$ C.L. contours which correspond to an upper limit of 2.3 signal events under background-free assumptions are superimposed. }

\item Fig.~\ref{fig:DM_kineDis_11} and Fig.~\ref{fig:DM_kineDis_12} show kinematic distributions for the process $e^{+} e^{-} \rightarrow \chi_2 \chi_1 \rightarrow (\chi_1 e^{+} e^{-} \chi_1)$ (dashed lines) and this process plus $e^{+} e^{-} \rightarrow h_{2} A^{\prime} \rightarrow (\chi_1 \chi_1) (\chi_2 \chi_1) \rightarrow (\chi_1 \chi_1) (\chi_1 e^{+} e^{-} \chi_1)$ (solid lines) based on BP-1 to BP-4 mentioned in the main text. The plots indicate that these two processes contribute differently to various kinematic distributions.
\end{itemize}

\section*{Acknowledgments}
PK is supported by KIAS Individual Grant No. PG021403.  
CTL and XQW are supported by the National Natural Science Foundation of China (NNSFC) under grant No.~12335005 and the Special funds for postdoctoral overseas recruitment, Ministry of Education of China. The authors gratefully acknowledge the valuable discussions and insights provided by the members of the China Collaboration of Precision
Testing and New Physics (CPTNP).

\end{document}